\documentstyle[graphicx,epsf]{basi}
\begin{document}
\title{Pre-main sequence stars, emission stars and recent
star formation in the Cygnus Region}
\author[Bhavya et al.]
{Bhavya B.$^{1,2}$, Blesson Mathew$^2$ \& Annapurni Subramaniam$^2$\\
$^1$ Cochin University of Science and Technology, Cochin\\
$^2$ Indian Institute of Astrophysics, Bangalore - 560034\\}
\maketitle
\label{firstpage}
\begin{abstract}
The recent star formation history in the Cygnus region is studied
using 5 clusters (IC 4996, NGC 6910,
Berkeley 87, Biurakan 2 and Berkeley 86). The optical data from the literature
are combined with the 2MASS data to identify the pre-main sequence (pre-MS)
stars as stars with near IR excess.
We identified 93 pre-MS stars and
9 stars with H$_\alpha$ emission spectra.
The identified pre-MS stars are used to estimate the turn-on age 
of the clusters. The duration of star formation was estimated as the
difference between the turn-on and
the turn-off age. We find that, NGC 6910 and IC 4996 have been forming stars
continuously for the last 6 -- 7 Myr, Berkeley 86 and Biurakan 2 for 5 Myr
and Berkeley 87 for the last 2 Myr.  This indicates that
the Cygnus region has been actively forming stars for the last 
7 Myr, depending on the location. 
9 emission line stars were identified in 4 clusters, using slit-less spectra
(Be 87 - 4 stars; Be 86 - 2 stars, NGC 6910 - 2 stars and IC 4996 - 1 star).
The individual spectra were obtained and analysed to estimate 
stellar as well as disk properties. 
All the emission stars are in the MS, well below the turn-off, in the core hydrogen 
burning phase. These stars are likely to be Classical Be (CBe) stars. Thus CBe phenomenon
can be found in very young MS stars which are just a few ( 2 -- 7) Myrs old.
This is an indication that CBe phenomenon need not be an evolutionary effect.
\end{abstract}



\section{Introduction}
\label{sec:intro}
Cygnus region, located between $70^o<l<80^o$ 
is a region of recent star formation activity in the 
Milky Way and is rich in massive early type stars concentrated in OB 
associations. The presence of nebulosity
and massive stars indicate that the stars have been forming till
very recently and the young clusters found here are the result of the recent
star formation event. Though the above fact is known, what is not known is that
when this star formation process started and how it proceeded in the region.
Thus the aim of the paper is to understand the star formation timescale
in this complex, using the recently formed young open clusters.

Though one assumes that all the stars in a cluster have the same age,
this assumption is not valid when the candidate cluster is very young.
In the case of young clusters ($\le$ 10 Myr), there is a chance
for a spread in the age of the stars, depending on the duration
of star formation. An estimation of this formation time-scale in the
clusters formed in a star forming complex, will indicate the duration
of star formation and its direction of propagation within the complex.
In principle, duration of star formation is defined as the 
difference between the
ages of the oldest and the youngest star formed in the cluster.
In practice, the age of the oldest star is assumed as the age of that star
which is about to turn-off from the main-sequence (MS) (turn-off age) and
the age of the youngest star is the age of the youngest pre-MS star (turn-on age).
The turn-off age of many clusters are known, but the turn-on age 
is not known for most of the clusters. The estimation of turn-on age
requires the identification of pre-MS stars in the cluster. 
Hence the duration of star formation in 
young star clusters is not clearly understood. 
This quantity cannot be
estimated for older clusters where all the stars have reached the MS.
The star clusters  studied here are
Berkeley 86, NGC 6910, IC 4996, Berkeley 87 and Biurakan 2.
We aim to combine the ages and the star
formation duration in these clusters to understand how the star formation 
proceeded in the Cygnus region. 

A number of stars are known to show H$_\alpha$ emission in this region.
The photometric analysis of the clusters suggested the possible presence of a few
emission line stars. As a followup, we searched, identified and studied these stars.
Since the estimation of cluster parameters will help in identifying the
ages and spectral types of these stars, their emission properties 
can be linked to their evolutionary status. 
Early type emission stars are broadly classified as Classical Be (CBe) stars and
Herbig Ae/Be (HAeBe) stars, which belong to intermediate mass pre-MS category.
CBe stars are fast rotators whose circumstellar disk is formed through
decretion mechanism (wind/outflow) (Porter \& Rivinius 2003). HAeBe stars are
found to possess a natal accretion disk which is a remnant of star formation
(Hillenbrand et al. 1992). The emission is found to come from this equatorial disk as
recombination radiation, mainly in Balmer lines like H$_\alpha$ and H$_\beta$.
We have used slitless spectroscopy to search for these emission stars. Emission stars
are found to show an IR excess, which is a combination of free-electron excess
and dust excess. We combine the parent cluster parameters and the spectral properties
of emission stars, and try to classify them. Fabregat \& Torrejon (2000)
suggested that emission stars found in clusters younger than 10 Myr belong to the
HBe type and CBe stars are found only in clusters older than 10 Myr. 
On the other hand, the HAeBe phenomenon is known to last
only for a few Myr (Hillenbrand et al. 1992). Thus
it will be interesting to identify and classify emission stars in clusters
younger than 10 Myr.

In the following section, we present the data used and discuss the analysis 
of the photometric data. Section 3 deals with the results obtained for the 5 clusters.
In section 4, we present the spectra  and discuss the emission line stars.
A discussion
of the star formation in Cygnus region is presented in section 5. The conclusions drawn from this
study are listed in section 6.

\section {Observations and Analysis}
The photometric data of clusters in the UBV 
optical pass bands are taken from WEBDA
(http://obswww.unige.ch/webda), a website devoted for star clusters.
The photometric data in the near-infrared (NIR) in J, H, K 
pass bands are taken from 2MASS data 
(http://www.ipac.caltech.edu/2mass/overview/access.html).
The reddening and extinction towards the clusters are already available in WEBDA.
The ZAMS data for distance determination were taken from Schmidt-Kaler (1982).
The pre-MS isochrones for estimating the pre-MS age of the cluster are 
taken from Siess et al. (2000)

\subsection{Identification of pre-MS stars} 
In the optical colour-magnitude diagram (CMD), both pre-MS stars and 
field stars occupy the same region (right side of the MS) and
we cannot distinguish between the two in the CMD.
Spectroscopic identification of pre-MS stars could be done at the
brighter end, but is inefficient for fainter stars. 
In the case of pre-MS stars, the dust grains
in the circumstellar disk absorb a fraction of visible or ultraviolet photons 
emitted by the central star, heating it to a higher temperature.
The disk re-radiates the absorbed energy in infra-red wavelengths. 
Thus the presence of NIR excess is used to distinguish between the pre-MS stars and 
field stars in the optical CMD. 
Thus, pre-MS stars can be identified using near-infrared (NIR) photometry. 
By cross-correlating optical and infrared images of the cluster field, 
the J, H, K magnitudes of optically identified stars were found. 
We used the 
(J$-$H) vs (H$-$K) NIR colour-colour diagram 
to identify stars with NIR excess and to determine their nature (Subramaniam et al. 2005).
The intrinsic location of un-reddened giants and MS stars,
and the normal reddening vector, 
which limits the reddening due to normal interstellar dust are used to distinguish
various types of stars with NIR excess. 
By relating their position in the NIR colour-colour diagram with the general
location of HAeBe and CBe stars, we identify candidate stars belonging to
the above classes.

\subsection{Estimation of turn-on age}    
The stellar magnitudes were corrected for reddening and extinction.
Though the distance to the clusters have already been estimated, we 
re-estimated the distance, which in most cases turned out to be similar to the
already estimated value.
The dereddened and extinction corrected CMD is fitted with ZAMS to estimate
the distance modulus (DM). 
The turn-off age of the clusters have already been estimated and we adopt these values.

In the NIR colour-colour diagram, stars which have NIR excess were identified.
These stars were plotted
on the optical CMD, as candidate pre-MS stars (indicated by a different symbol).
Pre-MS isochrones of different ages from Siess et al. (2000) 
were fitted to these stars on the CMD.
The presence of pre-MS stars in more than one isochrone shows 
that the pre-MS stars have a range in age, which in turn would indicate that 
the stars were formed over a time-scale.
The age of the youngest pre-MS star would be the turn-on age of the cluster.

\section{Results}

\subsection{IC 4996}
IC 4996 is located ($RA=20^h14^m24^s$, $Dec=+37^o29'$, $l=75.36^o,b=1.31^o$)
in the direction of the Cygnus and is part of a star forming region. 
An IRAS map of the region (Lozinskaya $\&$ Repin 1990) shows 
the presence of a dusty shell around the cluster. Zwintz and Weiss (2006) have performed 
time series CCD photometry in Johnson B and V filters to find 40 stars to lie in the 
classical instability strip of the 113 stars analysed in the cluster. They have discovered 
two \ delta Scuti-like pre-MS stars in the cluster. The parameters obtained by 
Delgado et.al (1998) for the cluster are $E(B-V)$=0.71 $\pm$ 0.08 mag, 
DM = 11.9 mag and age of 7.5 $\pm$ 3Myr. They suggested a 
number of pre-MS
stars in the cluster, which are located at 0.5 and 1 magnitude above the MS
in the V vs (B$-$V) CMD, around the location of spectral types A-F. 
Delgado et.al (1999) estimated the spectral types and heliocentric radial velocities 
for 16 stars in the cluster and the mean radial velocity was found to be
$-$12 $\pm$ 5 kms$^{-1}$.  
Vansevicius et.al (1996) estimated the distance and age of the cluster to be 1620 pc
and 9 Myr respectively, using BVRI CCD photometry. Pietrzynski (1996) performed variability 
studies on the cluster and found an RR Lyrae type variable 
and another eclipsing system. According to WEBDA, cluster contains two
B type emission stars.\\

We used the optical data of Delgado et al. (1998), available from WEBDA.
From the field of the cluster, 84 optically identified stars were 
found to have counterparts in the 2MASS data. We combined the
V, (B$-$V), (U$-$B) (taken from WEBDA) and  J, H, K (taken from 2MASS)
magnitudes for 84 stars.
From the NIR color-color diagram (figure 1), 22 members are found
to be located below the 
reddening vector. Among them, 15 could be considered 
to be candidate pre-MS stars and these are shown as dots with open circles
around them. Star numbered 23 is found to have H$_\alpha$ 
emission and shown as labeled filled circle. 
The MS and the giant star locations in the CMD are shown
(Bessell \& Brett 1988). The location of T-Tauri stars is shown as the dashed straight
line (Meyer et al. 1997). The location of Be stars is taken from Dougherty et al. (1994)
and the location HAeBe stars is taken from Hernandez et al. (2005).
The diagram on the left is not corrected for the reddening towards the cluster, whereas
the figure on the right is corrected for the reddening. 
The reddening value as estimated from the ZAMS fitting is used for dereddening.
The relation from Bessell \& Brett (1988)
is used for converting reddening in the optical to NIR. 
This could also arise due to differential reddening.
Star 23 is likely to be a CBe candidate.
 
The identified stars are reddening and extinction
corrected by taking E(B$-$V) as 0.7 mag and R$_v$ as 3.1.
The DM is estimated as 11.8 $\pm$ 0.2.
and the distance as 2291 $\pm$ 225 pc, which is similar to that estimated
by Delgado et al. (1998). The turn-off age of 
the cluster is 7.5 $\pm$ 3 Myr (Delgado et.al 1998).
Pre-MS isochrones for ages 0.25, 0.5, 1, 3, 5 and 7 Myrs are
plotted as shown in the Figure 2. 
The presence of pre-MS stars on or near all the pre-MS isochrones shows that
the stars have been forming continuously for the last 7 Myr.
Also, the presence of pre-MS stars near the tip of the MS indicates that
some of the high mass stars could be very young and the star formation stopped very recently.
Thus, the duration of star formation is similar to the age of the cluster, i.e.,
$\sim$ 7 Myr. The H$_\alpha$ emission star is located slightly to the
right of the MS  and there are a few candidate pre-MS stars located nearby.
Therefore, this star could either be a MS CBe star or a pre-MS HBe star. 
If it is a pre-MS
star, then the isochrones indicate its age to be between 0.5 -- 0.25 Myr.
If it is a CBe star, then its age is $\le$ 7 Myr.
 
To study the distribution of pre-MS stars within the cluster, we compared the location of
pre-MS stars and normal stars, as shown in Figure 3. 
Pre-MS stars are shown as filled circles and normal stars as open circles. The
emission line star (23) is also labeled.
The pre-MS stars are found to be located relatively closer to the center 
of the cluster when compared to the normal stars.

\subsection{NGC 6910}
NGC 6910 
($RA=20^h21^m18^s$, $Dec=40^o 37'$, $l=78^o .66, b=2^o .03$) 
is a young open cluster 
located in the Cygnus region and is a part of the Cygnus OB9 association. 
This cluster is located in the 
core of the star forming region, 2 Cygni. It is surrounded by a series
of gaseous emission nebulae which resemble Barnard's loop in the Orion. 
From UBV CCD observations down to V=18 mag for 206 stars in the 
cluster, Delgado et.al (2000) 
found eleven pre-MS stars of spectral type A to G.
They estimated the cluster parameters to be
$E{(B-V)}$ = 1.02 $\pm$ 0.13, DM = 11.2 $\pm$0.2 and 
age = 6.5 $\pm$ 3 Myrs. Kolaczkowski et al. (2004) found four beta Cep-type 
stars along with three 
H$_\alpha$ emission stars, while searching for variable stars in the
cluster. They  suggest a
possibility of finding a large number of beta Cep stars in 
the cluster due to 
higher metallicity of the cluster. Using $VI_c$ and $H_\alpha$ photometry they have 
determined an age of 6 $\pm$ 2 Myr, DM of 11.0 $\pm$ 0.3 mag and 
an $E(B-V)$ value varying from 1.0 to 1.4 magnitude.
Shevchenko et al. (1991) studied the cluster using UBVR photoelectric 
photometry. From the photometry of 132 stars,
they found the HAeBe stars BD +40 41 24 and BD +41 37 31 to be associated with it. 
The extinction is 
high in the region with a value of $E(B-V)$ = 1.2 mag and the value of 
R = 3.42 $\pm$ 0.09.  
Using intermediate-band photoelectric photometry for 
16 cluster members, Crawford et.al (1977)
obtained a reddening value of $E{(b-y)}$=0.75 mag and a DM of 10.5 mag.
They found that a type Ia super giant star
of $M_v$ = -6.9 to be associated with the cluster. According to WEBDA there is no 
emission star in the cluster.\\

The U, B$-$V, U$-$B values of 148 stars are taken from WEBDA (Delgado et al. 2000).
The J, H, K values of these stars are taken from 2MASS by cross-correlating
optical and IR images. NIR colour-colour diagram is plotted as shown in Figure 4.
47 stars were found to be located below the reddening vector.
Out of them, 23 members were found to have relatively large IR excess and were considered as
candidate pre-MS stars. Stars, 26 and 181 do not have CCD photometry, we have
used the UBV photographic photometry of Hoag et al. (1961). These stars show
$H_\alpha$ emission and are shown as dark filled circles.  After dereddening,
star 181 can be found very close to the CBe location, indicating its CBe nature. 
 
The reddening and extinction 
corrections are done by assuming E(B$-$V) as 0.97 mag and $R_v$ as 3.1. 
ZAMS fitting is done as shown in Figure 5, the DM is estimated as
11.8$\pm$0.2 mag and the distance as 2291$\pm$225 pc. The 
turn-off age of the cluster is 6.5$\pm$3 Myr (Delgado et al. 2000).
The pre-MS isochrones for 0.5, 1, 2, 3, 5, 7.5 and 10 Myrs are overlaid
on the CMD.
The presence of pre-MS stars 
with these ages shows that star formation processes are occurring
in the cluster continuously. More number of stars are formed
in the early stages and continuing up to 3 Myr. 
After that the star formation seems to have continued
 but in a slower rate, till 0.5 Myr. We find that a few pre-MS stars are as old as
10 Myr. Also for this cluster, we get the duration of star formation
as the cluster age itself $\sim$ 7 Myr, though there is an indication that
the star formation started 10 Myr ago. 
The H$_\alpha$ emission star 181 is found to be 
located near the 0.5 Myr isochrone, indicating that this could also be a candidate
HBe star. The star 26 is located to the left of the MS, hence we do not discuss
the nature of this star based on the photometry. The peculiar location may be due to
two reasons - we used the photographic data for these two stars and there
may be some error in the data of this star, or, the reddening is very different for this star.
If these two stars belong to the CBe class, then these stars are $\le$ 7 Myr old.
 
The pre-MS
stars are shown separately as dark circles and normal 
stars are shown as open circles, as
in Figure 6. We find that the pre-MS stars are scattered throughout the face
of the cluster. The emission stars 26 and 181 are located very close to each other.
         
\subsection{Berkeley 87}   

       The open cluster, Berkeley 87 
(Dolidze 7,$ RA=20^h21^m42^s, Dec=37^o.22$),
 located at $l=75^o .71, b=+0^o .31$ is a
 sparse grouping of early-type stars lying in a heavily
 obscured region in the Cygnus. Turner \& Forbes (1982)
derived a distance of 946 $\pm$ 26 pc and an age of 1-2 Myr 
for the cluster from UBV photometry of 105 stars. The age as
reported in WEBDA is 14 Myr.
The distance was estimated to be 0.9 kpc from the interstellar 
line depth of the cluster members (Polcaro et al. 1989). 
It is part of the star formation region ON 2, which harbours many 
compact HII regions, strong OH masers and CO and ammonia molecular clouds. 
Diffuse emission in CIV 5802-12 A doublet and X-ray emission in 2-6 keV 
range has been detected in the cluster, 
which is interpreted as due to the interaction of the strong wind from a  
Wolf Rayet star (ST 3) of 
velocity 5200 kms$^{-1}$ with the cluster members. A high energy
 Gamma ray source 2CG 075+00 ($\le$100 Mev) 
has been found to be associated with this cluster. Be 87 contains a 
peculiar emission-line B super giant HDE 229059 (Hiltner 1956), 
two faint objects (VES 203 and VES 204) which show 
H$_\alpha$ in emission (Coyne et.al 1975), the red super giant 
BD +37 39 03 and the faint 
variable star V439 Cygni. According to WEBDA the cluster contains 5 
B type emission stars.\\

The optical data of 94 stars (Turner \& Forbes 1982) were combined with
the 2MASS data.
 19 stars were identified to have possible NIR excess from the NIR colour-colour
diagram (figure 7). Out of them, 13 members having relatively 
 higher (H$-$K) magnitudes were grouped as candidate pre-MS stars. Stars 9, 15, 38 
 and 68 (Turner \& Forbes 1982) are stars with H$_\alpha$ emission, 
shown as filled circles.
 Star 38 and 15 occupy the HAeBe location, after the reddening correction. Stars
9 and 68 are likely to be CBe stars. There is significant differential
reddening in this cluster, as discussed below. Therefore, the above classification
should be taken with some caution.
 
 In the NIR colour-colour diagram, many stars are found outside the
 reddening vector. Some stars are more reddened than others
 i.e., there is a variation of  reddening in different parts of
 cluster field. All cluster members do not have the same $E(B-V)$ value.
 Color correction and extinction correction are done by choosing
$E(B-V)$ as 0.45 for stars
 having (H$-$K) value less than 0.1 and as 1.35 for stars having (H$-$K)
 value higher than 0.1 along with $R_v$=3.1. The variation in $E(B-V)$
 gives an indication of the non-uniform distribution of interstellar material
within the cluster. The resulting CMD is shown in figure 8.
The wide MS in the CMD is an indication of significant differential reddening
across the cluster.
The dereddening in the NIR is done using the lower value of the two reddening 
values.

 ZAMS fitting estimated the DM
 to be 10.8 $\pm$ 0.2 mag. The distance estimated is 1445 $\pm$ 145 pc, higher
than the value estimated by Turner \& Forbes (1982). 
 The pre-MS isochrones for the ages 0.15, 0.2, 0.25, 0.5, 1, 1.5 and 2 Myrs are
 plotted on the CMD (shown in Figure 8). Most of the pre-MS stars are located 
 in isochrones of ages ranging from 0.15 to 2 Myr. Therefore, the turn-on
age of the cluster is less than 1 Myr, which is consistent with the
recent star formation activity in the cluster vicinity. This recent star
formation activity is likely to have started 2 Myr back, as indicated by the
age of the pre-MS stars.
 The cluster has a turn-off age of 14 Myr, as indicated in WEBDA. On the
other hand, the age of 2 Myr as estimated by Turner \& Forbes (1982), is in good agreement
with our estimation of the starting time of the star formation. Therefore, this may be a  very young
cluster with a turn off age of 2 Myr. Therefore, we consider the turn-off age
of Be 87 as 2 Myr. The emission line stars in this cluster are very young.
If they are HAeBe stars, then they are as young as 0.15 Myr. If they
are CBe stars, then they are $\le$ 2 Myr old. Thus, if these stars are CBe stars,
these might be one of the very young CBe stars known.

The location of pre-MS stars in the field of the cluster is shown in figure 9.
It can be seen that there is a clustering
of pre-MS stars as well as the emission stars towards 
the central region of the cluster.  
   
\subsection{Biurakan 2}
       The galactic cluster, Biurakan 2, is 
 ($RA=20^h09^m12^s,
 Dec=35^o.29$,  $l=72^o.751, b=1^o.345$) also located 
 in the vicinity of the association, Cygnus OB-1.
The cluster was studied by Dupuy and Zukauskas (1976) 
using photoelectric and photographic photometry.
They estimated 
an average value of E(B$-$V)=1.39$\pm$.09 towards the cluster, 
DM as 10.8$\pm$0.2 mag, which corresponds to a distance of 1445$\pm$133 pc 
and age $\le$10 Myr. According to WEBDA there is no Be star in the cluster.\\
 
By cross-correlating the optical (Dupuy \& Zukauskas 1976) and NIR data of
the cluster field, 114 stars were identified to have both data.
NIR colour-colour diagram is plotted as shown in Figure 10.
 33 stars were found to be
 located below the reddening vector and 15 stars having 
 higher (H$-$K) were grouped as candidate pre-MS stars. No star with 
 H$_\alpha$ emission is observed in this cluster. 
  
 Assuming $E{(B-V)}$ as 0.45 mag and $R_v$ as 3.1, colour  correction
 and extinction correction were done and the resulting CMD 
 is shown in Figure 11.
We re-estimated the DM as  11.2 $\pm$ 0.2 mag. The distance 
obtained is 1737.8 $\pm$ 175 pc.
 The pre-MS isochrones for the ages 0.5, 1, 1.5, 2, 3 and 5 Myrs are plotted.
 The presence of pre-MS stars in these isochrones shows that the pre-MS stars
have an age range of 0.5-5 Myr. Most of the pre-MS stars are younger than 2 Myr 
and only four stars were found to be older than 5 Myr.
This indicates that the star formation started 5 Myr back, but was very active for the
last 2 Myr.
   The turn-off age of the cluster obtained by the previous study 
is 10 Myr  or less (Dupuy et.al 1976). The turn-off age estimation is difficult and
inaccurate due to the absence of evolved stars. It is quite likely that the turn-off age
of the cluster is 5 Myr, which is the time when the star formation started. 
The turn on age is less than 1 Myr. Thus the duration of star formation in this cluster
is about 5 Myr.

 We compared the location of pre-MS stars in the cluster field, as shown
in figure 12. It can be seen that the pre-MS stars are distributed across
the cluster field and do not show any central concentration.

 \subsection{Berkeley 86}
Berkeley 86, a young open cluster, is one of the three nuclei of 
the OB association Cyg OB 1.
The core of this cluster is located at $ l=76^o.7, b=1^o.3 $.
Optical photometry was carried out in UBVRI bands by Deeg \& Ninkov (1996) and 
Massey et al. (1995). 
They found the cluster to be a young one with a turn-off age of 5 Myr, reddening of 1 mag 
and an initial 
mass function close to the Salpeter value. Be 86 hosts the famous eclipsing binary 
system V444 Cygni (Forbes 1991).
Stromgren photometry was obtained by Delgado et al. (1997) down to V=19 mag, from which 
they estimated an age of 8.5 Myr and a distance of 1659 pc. 
Forbes (1981) found Be 86 to lie at a distance of 1.72$\pm$0.20 kpc with a 
reddening of E(B$-$V)=0.96$\pm$0.07 mag.
The cluster is located in the Orion spiral feature and is a possible member 
of the Cygnus OB 1 association at 1.8 kpc. 
An age of 6 Myr was estimated based on the earliest spectral type of O9.
Vallenari et al. (1999) studied about 2000 stars in the
field of Berkeley 86 down to K $\sim$ 16.5 mag 
using near-infrared photometry in J and K bands. They have found a 
number of pre-MS stars from 
(V$-$I) vs (I$-$K) plot and J vs (V$-$J) diagram. They do not estimate the
ages of their candidate pre-MS stars. According to WEBDA cluster contains 
3 Be stars.\\

From the field of Be 86 about 191 stars are identified with 
V, (B$-$V), (U$-$B), J, H, K magnitudes. Optical data is taken from
WEBDA (Deeg \& Ninkov 1996). NIR colour-colour diagram is 
plotted for the identified stars as shown in Figure 13.
36 stars are 
found to be located below the reddening vector, showing 
IR excess. 27 stars are found to have more 
IR excess (large (H$-$K) magnitude) and could be
candidate pre-MS stars. Three more stars are located
below the reddening vector (not shown in the figure) with
very large (H$-$K) value. These stars are located in HAeBe
location, could be probable HAeBe stars. 
The stars 9 and 26 (Forbes 1981) for which H$_\alpha$ emission was
observed are shown as filled and labeled circles. In the reddening
corrected figure (right panel), these stars occupy the CBe location.

For the identified stars reddening and extinction 
correction are done by assuming $E(B-V)$ as 0.95 mag and $R_v$ as 3.1.
The resulting CMD is shown in Figure 14.
We re-estimated the distance as 1585$\pm$160pc.
 In the CMD,  many stars fainter than $V_0$ = 14 mag and having 
IR excess are located to the left of ZAMS. These may be field stars or stars with
different reddening.
 The turn-off age of the cluster is 6 Myr (Forbes et al. 1992). 
 Pre-MS isochrones of ages 0.25, 0.5, 5, 7.5 and 10 Myrs are
 shown. The pre-MS stars are distributed in two age groups.
The younger group is found to be younger than 1 Myr, while the older group
is as old as or older than 5 -- 7 Myr. It is quite likely that the star formation
in this region started around 5 Myr and had a relatively low level star formation
till 1 Myr. The cluster region has experienced vigorous star formation in the last 1 Myr. 
The duration of star formation is about 5 Myr, which is similar to the cluster, Biurakan 2.
Star 9 lies on the 0.25 Myr pre-MS isochrone and 26 is located on the 0.5 Myr isochrone. 
Thus these stars are could be younger than 1 Myr, which makes them as candidate HAeBe stars.
Otherwise, these stars could be 5 Myr old or slightly older, which would make them as
candidate CBe stars.
The Pre-MS stars are found to be distributed throughout the cluster field, as shown in Figure 15. 
\section{Spectroscopic detection of emission line stars}

\begin{table}
\caption{Log of observations of Individual Clusters in the slit-less mode}
\vspace{0.5cm}
\centering
\begin{tabular}{ccccccc}
\hline
Date of & Cluster & Filter/ & Exposure & RA & Dec\\
observation &    & Grism  & Time   & hh:mm:ss & deg:mm:ss\\
\hline
30-10-2003 & IC 4996 & R & 5 & 20:16:30 & +37:38:00 \\
           &         & R/Gr5 & 60 & & \\       
           &         & R/Gr5 & 300 & & \\  
22-11-2004 & NGC 6910 & R & 5 & 20:23:12 & +40:46:42 \\ 
           &          & R/Gr5 & 60 & & \\   
           &          & R/Gr5 & 180 & & \\
           & NGC 6910 & R & 10 & & \\  
           &          & R/Gr5 & 600 & & \\
15-07-2005 & NGC 6910 & R & 10 & & \\
           &          & R/Gr5 & 600 & & \\ 
22-11-2004 & Berkeley 87 & R & 5 & 20:21:42 & +37:22:00 \\ 
           &             & R/Gr5 & 120 & & \\
           &             & R/Gr5 & 480 & & \\
15-07-2005 & Berkeley 87 & R & 10 & & \\ 
           &             & R/Gr5 & 900 & & \\
21-07-2006 & Berkeley 87 & R & 10 & & \\ 
           &             & R/Gr5 & 900 & & \\
27-06-2005 & Berkeley 86 & R & 10 & 20:20:24 & +38:42:00 \\ 
           &             & R/Gr5 & 600 & & \\ 
08-06-2004 & Berkeley 86 & R & 30 & & \\           
           &             & R/Gr5 & 300 & & \\
           &             & R/Gr5 & 600 & & \\
\hline
\end{tabular}
\end{table}

\begin{table}
\caption{Log of observations of Emission Stars in Clusters (Slit Spectra)}
\vspace{0.5cm}
\centering
\begin{tabular}{ccccccc}
\hline
Date of & Cluster & Star & Grism/ & Exposure & RA & Dec\\
observation &     &      & Slit  & Time   & hh:mm:ss & deg:min:ss\\
\hline
15-06-2005 & IC 4996     & 23 & Gr7/167$\mu$ & 600 & 20:16:29.05 & +37:38:53.9\\                      
           &             &    & Gr8/167$\mu$ & 600 & & \\            
29-07-2005 & NGC 6910    & 26 & Gr7/167$\mu$ & 900 & 20:23:09.76 & +40:45:53.7 \\            
           &             &    & Gr8/167$\mu$ & 900 & & \\                                
           &             & 181 & Gr7/167$\mu$ & 900 & 20:23:11.82 & +40:43:28.8 \\ 
           &             &     & Gr8/167$\mu$ & 900 & & \\                       
09-10-2005 & Berkeley 87 & 9 & Gr7/167$\mu$ & 600 & 20:21:29.72 & +37:26:25.1 \\ 
           &             &   & Gr8/167$\mu$ & 600 & & \\
           &             & 15 & Gr7/167$\mu$ & 900 & 20:21:33.59 & +37:24:51.7 \\ 
           &             &    & Gr8/167$\mu$ & 900 & & \\
           &             & 38 & Gr7/167$\mu$ & 900 & 20:21:59.9 & +37:26:25.1 \\ 
           &             &    & Gr8/167$\mu$ & 900 & & \\           
           &             & 68 & Gr7/167$\mu$ & 900 & 20:21:28.9 & +37:26:19.9 \\               
           &             &    & Gr8/167$\mu$ & 900 & & \\ 
27-06-2005 & Berkeley 86 & 9 & Gr7/167$\mu$ & 900 & 20:20:10.55 & +38:37:29.8 \\         
           &             &   & Gr8/167$\mu$ & 900 & & \\ 
           &             & 26 & Gr7/167$\mu$ & 900 & 20:20:22.78 & +38:38:12.3 \\
           &             &    & Gr8/167$\mu$ & 900 & & \\           
\hline
\end{tabular}
\end{table}

The spectroscopic observations of the emission stars 
in clusters have been obtained using  the HFOSC instrument available 
with the 2.0m Himalayan Chandra Telescope, located at
HANLE and operated by the Indian Institute of Astrophysics.  
Details of the telescope and the instrument are available 
at the institute's homepage
(http://www.iia.res.in/). We have used slitless spectroscopic 
technique to identify emission stars in a cluster 
(Subramaniam et al. 2005). The log of observations of 
individual clusters in the slit-less mode is tabulated in Table 1.
Slit spectra for identified emission
stars were taken using Grism 7 (3800\AA-6800\AA) and 
167 $\mu$ slit combination in the blue region which 
gives an effective resolution of 1330. The spectra in the
red region were taken using Grism 8 (5800\AA-8850\AA) 167 $\mu$ 
slit setup, which gave an effective resolution of 2190. 
The CCD used was a 2 K $\times$ 4 K CCD, where 
the central 500 $\times$ 3500 pixels were used for
slit spectroscopy. The log of observations of the identified 
emission stars in the slit mode is tabulated in Table 2.
All the observed spectra were wavelength 
calibrated and corrected for instrument sensitivity using 
IRAF tasks. The spectrophotometric standard, BD
+284211, was observed on corresponding nights and was
used for flux calibration. The resulting flux 
calibrated spectra were normalised and continuum fitted
using IRAF tasks. The spectra are shown in figures 16 (blue region)
and 17 (red region). Figure 18 shows the spectra of 4 stars which
show prominent lines in the red end of the optical spectrum.
The spectral
properties of stars in four clusters are discussed below.

NaI lines (5890, 5896\AA), Telluric $O_2$ bands (6867, 7600\AA) and 
TiO molecular band (6080-6390\AA) can be seen the spectra of 
emission stars in all the clusters.\\

{\bf IC 4996:}
The $H_\alpha$ profile of star 23 (GSC 03151-00898) shows emission in absorption feature 
and has OI(8446 \AA) line in emission.\\

{\bf NGC 6910:}
Star 181 has H$_\alpha$ 
and OI (8446\AA) lines in emission.
The H$_\alpha$ profile is found to show an asymmetry in profile
with a definite V/R ratio.
Star numbered 26 has an emission in absorption feature for H$_\beta$ 
and H$_\delta$ profiles.   

{\bf Be 87:}
Star 9 (B0.5Ve; GSC 02684-00007) shows double peaked 
emission for H$_\alpha$ profile
along with H$_\gamma$ and Paschen 14 line in emission. Spectral lines 
like OI (8446\AA), Calcium triplet (8498, 8542, 8662\AA) and FeII line 
(6384\AA) are seen in emission.
Star 15 has an equivalent width of -3.5 and -32.89 for 
H$_\beta$ and H$_\alpha$ profiles respectively, which is the highest among 
these 9 emission stars. Intense emission can be seen for 
spectral lines like OI (8446, 7771\AA), Paschen lines (P14(8598\AA), P12 (8750\AA)),
 Calcium triplet(8498, 8542, 8662\AA) and FeII lines 
(6384, 5316, 7712\AA).
For star 38, (B2IIIe; GSC 02684-00004) only  H$_\alpha$ 
is seen in emission among the
group of Balmer lines. Faint emission features of OI, FeII(4351, 4620\AA) 
and Calcium triplet are seen. He lines (4922, 4471, 4026, 5876, 6678\AA) 
are seen in absorption.
For star numbered 68, the H$_\alpha$ profile is seen in double 
peaked emission while H$_\beta$ 
is seen in absorption. The spectrum is more or less featureless with 
tentative emission in OI (7772, 8446\AA), Calcium and Paschen lines.\\

{\bf Be 86:}
For the star 9, the H$_\alpha$ profile is found to show a double peak profile with 
the red region intense over violet. Balmer lines other than  H$_\alpha$ 
are in absorption. He lines (4922, 4471, 4026, 5876, 6678\AA) 
are seen in absorption.
For the star numbered 26 intense emission can be seen for 
spectral lines like OI (7254, 8446\AA), 
Paschen lines (P14 (8598\AA), P12 (8750\AA)),
Calcium triplet (8498, 8542, 8662\AA) and FeII lines 
(6384, 5316, 7712\AA). 
P15 (8545\AA), CaII (8542\AA) lines and P13 (8665\AA), 
CaII(8662\AA) are blended due to the low resolution of the 
spectrum. In the blue region (3700-4500\AA), Balmer lines other than H$_\alpha$ and 
H$_\beta$ (core emission) are seen in absorption.

\subsection{Estimation of stellar and disk properties}
The equivalent width and FWHM of the spectral lines 
are estimated from a Gaussian fit, using routines in IRAF.
The FWHM of H$_\alpha$ is corrected for instrument 
line width using comparison lines. The rotation 
velocity has been calculated using the formula,
$$VSin{\it i}=c*FWHM/(2\lambda_0*(ln2)^{(1/2)})$$

 where 
c is the velocity of light, $\lambda_0$ is the 
central wavelength of the spectral line.
The rotation velocity estimated using H$_\alpha$  
emission line profile gives the rotation velocity 
of the disk while that from HeI (4471 \AA) absorption line profile 
gives the stellar rotation velocity.
The estimated parameters are tabulated in table 3. Many of the
previous estimates of the spectral types indicate  some
of the emission stars to be giants. We have tried to classify the
stars according to their photometry. The estimated spectral types
are given in table 3. It is evident from the optical CMDs that all the
emission stars are located either on the MS or very close to the  MS.
Also they are not located near the turn-off of the MS, but well below
the tip of the MS. This indicates that
none of the stars are evolving to the giant phase from the MS. Thus, 
we consider the luminosity class to be V for all the emission stars.

\begin{table}
\caption{Estimated parameters of 9 emission stars. 
The age estimated assuming them
to be pre-MS stars is tabulated in column 3, while the age, assuming them to be MS stars
is tabulated in column 4. The Spectral type is estimated from
the photometry (column 5). The H$_\alpha$ EW is in column 6. The stellar rotation velocity is
estimated from the FWHM of He I (4471 \AA) and disk rotation velocity from H$_\alpha$ are in
columns 7 and 8.
 }
\vspace{0.5cm}
\centering
\begin{tabular}{cccccccc}
\hline
Cluster &  Star &  Age & Age &   Sp. type &
 Eq.width&
 Disk Velocity & Star Velocity \\
 & & (pre-MS)& (MS)& & & kms$^{-1}$ & kms$^{-1}$\\
\hline
 IC 4996 & 23 & 0.5 &7& B2.5 & $-$18.99$\pm$1.13 & 268$\pm$6 & 301$\pm$5 \\
 NGC 6910& 26 & - &7& B2.5 &$ -$9.24$\pm$0.45 & 228$\pm$5 &  234$\pm$6 \\
  & 181 & 0.5 &7& B5 & $-$29.25$\pm$2.09 & 339$\pm$36 & 224$\pm$3 \\
 Be 87 & 9 & 0.2 &2& B1.5 & $-$22.87$\pm$1.22 & 352$\pm$8 & 266$\pm$7 \\
 & 15 & 0.1 &2& B1.5 & $-$32.89$\pm$1.87 & 196$\pm$3 & 264$\pm$3 \\
 & 38 & 0.2 &2& B1.5 & $-$7.29$\pm$0.46 & 199$\pm$7 & 229$\pm$5 \\
 & 68 & 0.5 &2& B9 & $-$6.84$\pm$0.23 & 353$\pm$7 & 99$\pm$5 \\
 Be 86 & 9 & 0.25 &6& B2.5 & $-$4.83$\pm$0.14 & 367$\pm$1 & 383$\pm$3 \\
 & 26 & 0.5 &6& B4 & $-$19.02$\pm$1.1 & 151$\pm$6 & 491$\pm$5 \\
\hline
\end{tabular}
\end{table}

The estimated rotational velocity of the star, as well as the disk is 
basically the Vsin{\it i}, which is the projected component. 
In general, we expect the disk to trail behind the star,
but we can see that in certain cases the rotational velocity of the disk 
is higher than that of the star. 
The stars 181 (NGC 6910), 9 (Be 87) and 68 (Be 87) have disk
velocity higher than the star. All the three stars have either
double-peaked H$_\alpha$ or asymmetric profile.
In principle, one has to de-convolve the peaks and then estimate the individual FWHM.
Since the spectral resolution is poor, we are unable to deconvolve the profile. Thus,
we have over estimated the disk velocity. Thus, excluding these three cases, the
rest of the stars show faster rotation, when compared to the disk.

It will be interesting
to study the relative rotational velocity between the disk and the star. Since the
projection angle will contribute identically to both the velocities, the relative
velocity is assumed to be independent of the projection angle {\it i}.
This relative velocity could
be a function of spectral type and hence we consider 5 stars which are of
similar spectral type. The stars considered are 
the emission stars Be 87(38), NGC 6910 (26), Be 87(15),
IC 4996(23) and Be 86(9). We have estimated
the difference in rotation velocity between the star and the disk and
it is plotted against the rotation velocity of the star (figure 19).
The stellar rotational velocity 
is found to be between 200 -- 400 kms$^{-1}$. The two stars in Be 86 have
relatively large velocities (one is of later spectral type). 
The disk is found to lag behind the
star by 0 - 40 kms$^{-1}$ for four stars, For one star, the disk is slower by
68 kms$^{-1}$. This star happens to have the highest value of equivalent width for H$_\alpha$,
which indicates a heavier disk. 

We find that out of the 9 emission stars, most are likely to be CBe stars.
There may only be at the most one-two HBe candidates, if at all.
Also, the rotational velocity of the emission stars are in the range of those of
the CBe stars, supporting their CBe nature.
We see that the CBe phenomenon starts very early
in the evolution of these stars, by about 2 -- 7 Myr. Since these stars 
are well below the turn-off
of the cluster, the CBe phenomenon starts much before the star evolves from the MS.

Capilla et al. (2000) found that younger clusters (age $\le$ 10 Myr) do not contain Be stars.
Clusters in the age interval 10-30 Myr are rich in Be stars. Almost all of them are
of spectral types earlier than B5, while late-type Be stars are scarce. These results
point towards an evolutionary interpretation of Be phenomenon, in the sense that Be stars
are close to the end of their MS life time. Capilla et al. (2000)
did not find any Be stars in NGC 6910.
The clusters studied here all have a turn-off age less than 10 Myrs and they are found
to have 9 emission stars which are likely to be CBe candidates. 2 stars are of
spectral type B5 and one of B7, which indicates the possibility of late type Be stars
present in young clusters which in turn means that the CBe phenomenon need not be an
evolutionary effect. Fabregat (2007) suggested that CBe phenomenon is an evolutionary effect
appearing at the second half of the life in the MS of a B star.
But this study suggests that CBe phenomenon can happen during the core burning
initial phase in the MS life time of the star.
\begin{table}
\caption{The estimated duration of star formation in the five clusters.}
\vspace{0.5cm}
\centering
\begin{tabular}{cccc}
\hline
Cluster & Turn-on & Turn-off & Duration of \\  
        & age (Myr) & age (Myr) &  star formation (Myr) \\ 
\hline
 IC 4996 & 0.5 & 7.5$\pm$3 & 7(continuous) \\
 NGC 6910 & 0.5 & 6.5$\pm$3 & 7(continuous) \\
 Berkeley 87 & 0.2 & $\sim$2 & 2 (continuous) \\
 Biurakan 2 & 0.5 & $\sim$6 & 5 (continuous/recurrent)\\
 Berkeley 86 & 0.2 & $\sim$6 & 6(continuous/recurrent)  \\
 
\hline

\end{tabular}
\end{table} 
\section{Location of clusters and direction of star
 formation in the Cygnus Region}

In the Cygnus OB region, 5 clusters are 
studied. The estimation of the turn-off age and turn-on
 age of the clusters along with their distance provides a
reasonable interpretation about the direction of 
propagation of star formation in the Cygnus OB
star forming complex.
     The longitude and distance to the cluster is used to derive
their location in the Cartesian Coordinate system as shown in Figure 20.
The estimated duration of star formation in the clusters studied are tabulated in table 5.
The location of clusters along with their age is used to understand
 the star formation.
This region, located between the galactic longitudes $l = 70^o - 80^o $,
has eight more open clusters, for which reliable distance and age estimates are available.
We include these clusters also, as shown in figure 21, to understand the star formation.
The clusters included are Collinder 419, Collinder 421, Ruprecht 172, NGC 6913,
NGC 6871,  Biurakan 1 and ASCC 111. The first four clusters are
located closer to us, closer than 1200 pc and the rest of the three clusters are
located very close to Be 86, Be 87 and Biurakan 2 . The ages as given in WEBDA for these
three clusters are 9 Myr, 18 Myr and 11 Myr, respectively. 
 
We interpret this figure keeping in mind that there may be significant error
associated with the estimated distance. The clusters can be grouped into
three groups, the first one is located 
around 1 kpc, the second group is located at $\sim$ 1500 pc and the third group, beyond
2000 pc. The farthest group consists of two clusters, NGC 6910 and IC 4996.
Both the clusters seem to have similar star formation history.
The second group seems to indicate a vigorous star formation, where the relatively
older clusters, Biurakan 2,
NGC 6871, Biurakan 1 and ASCC 111 
could have formed first, then Berkeley 86 and very recently, Berkeley 87.
In both the groups, the duration of star formation is about 5 -- 7 Myr. We estimate
a short duration for the star formation, 2 Myr, in the case of Berkeley 87. The three clusters
studied here indicate an enhanced star formation in the last 1 -- 2 Myr.  
Also, the major star
forming region with 6 clusters is located at $\sim$ 1500 pc, and this could be considered as
the most active star forming region within the longitude range $l = 70^o - 80^o$ and located at 
$\sim$ 1500 pc. No preferential location for the young pre-MS stars were found in the clusters.
The emission line stars were found in 4 clusters and one was found to be devoid of such stars.
Therefore, all clusters in a star forming complex may not harbour emission line stars. 
Thus, the recent star formation observed in the
Cygnus was started about 2 -- 7 Myr ago, depending on the location. This implies that
the star formation in large complexes can last up to 7 Myr.

\section{Conclusion}
The main conclusions of the present study can be summarised as follows:
\begin{description}
\item In the Cygnus region, 93 candidate pre-MS stars 
and 9 stars with H$_\alpha$ emission spectra are identified in 5 clusters.
\item The duration of star formation (estimated as the difference between the turn-off and
turn-on age) is estimated for five clusters. This indicates that the star formation
in the clusters were going on for 2 -- 7 Myr. The recent star formation in the Cygnus
region started $\sim$ 7 Myr ago.
\item CBe phenomenon
can be found in very young MS stars which are just a few ( 2 -- 7) Myrs old.
This indicates that the CBe phenomenon need not be an evolutionary effect.

\section*{Acknowledgements}
This research has made use of the WEBDA database, operated at the Institute for
Astronomy of the University of Vienna. This publication makes use of data products
from the Two Micron All Sky Survey.
Bhavya B., would like to thank Indian Institute
of Astrophysics for supporting her for MSc. project and for her ongoing thesis project.

\end{description}

\begin{figure*}
\epsfxsize=17truecm
\centerline{\epsffile{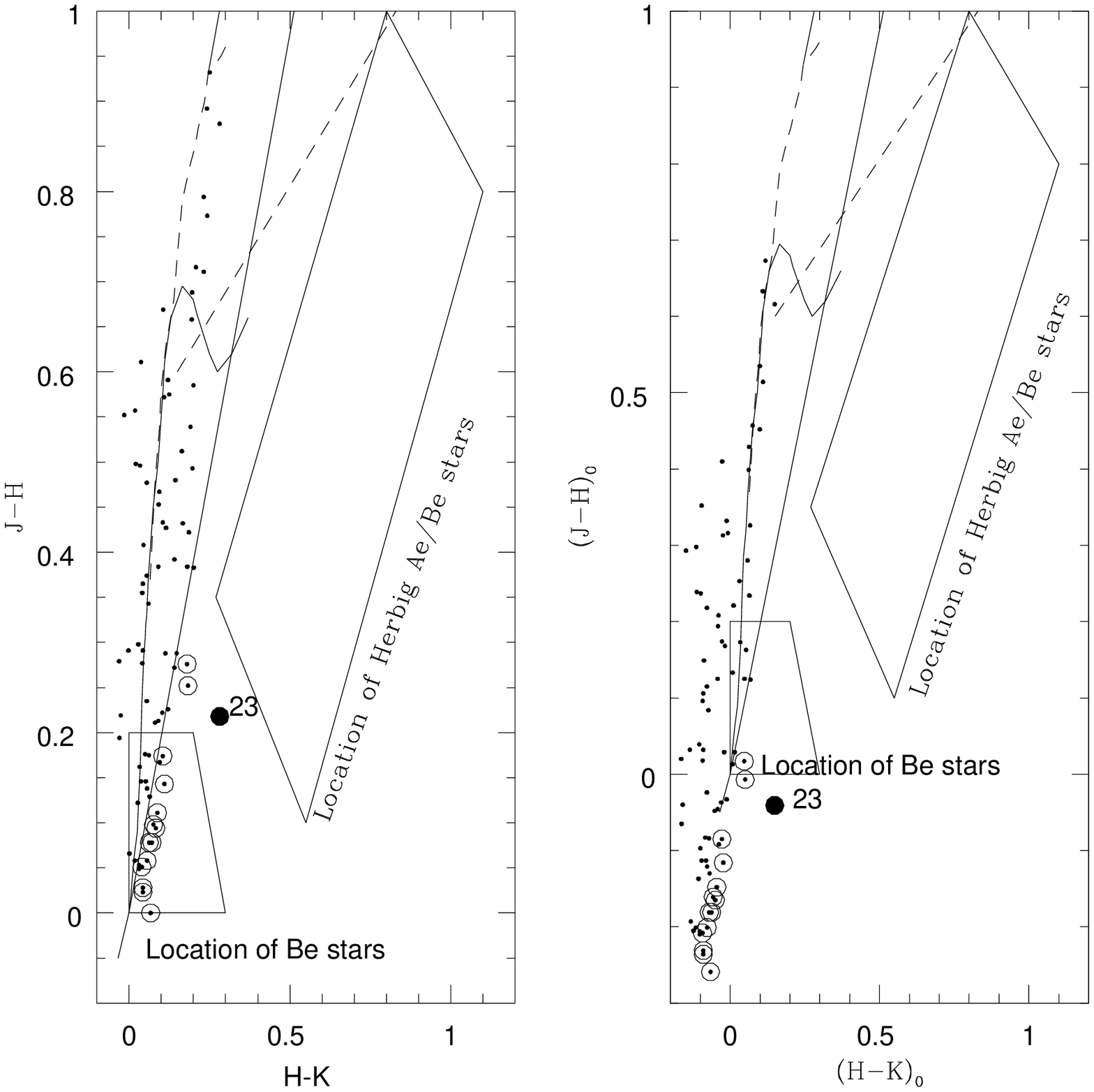}}
\caption{NIR colour-colour diagram for the cluster IC 4996. 
Dots indicate normal stars. Encircled dots indicate pre-MS stars. Stars 
with H$_\alpha$ emission are shown as labeled and filled circle. The left panel
shows colours uncorrected for reddening, whereas the right panel shows colours corrected
for reddening. The reddening value as estimated from the ZAMS fitting is used for NIR
dereddening, using the relation in Bessell \& Brett (1988).}
\end{figure*}
\newpage
\begin{figure*}
\epsfxsize=17truecm
\centerline{\epsffile{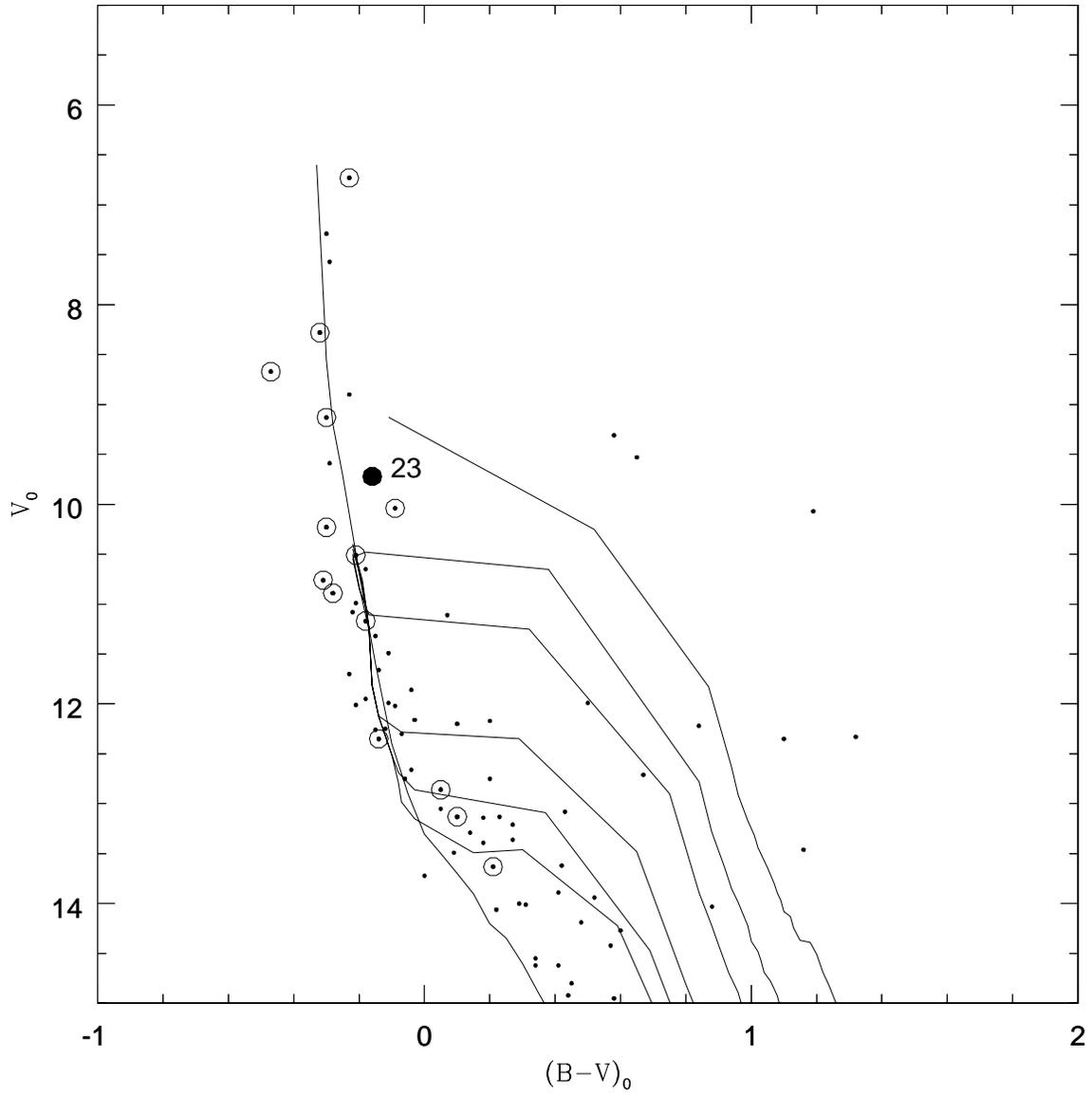}}
\caption{Pre-MS isochrone fittings for the cluster IC 4996. The isochrones for the ages
0.25, 0.5, 1, 3, 5 and 7 Myr are shown.}
\end{figure*}
\newpage
\begin{figure*}
\epsfxsize=17truecm
\centerline{\epsffile{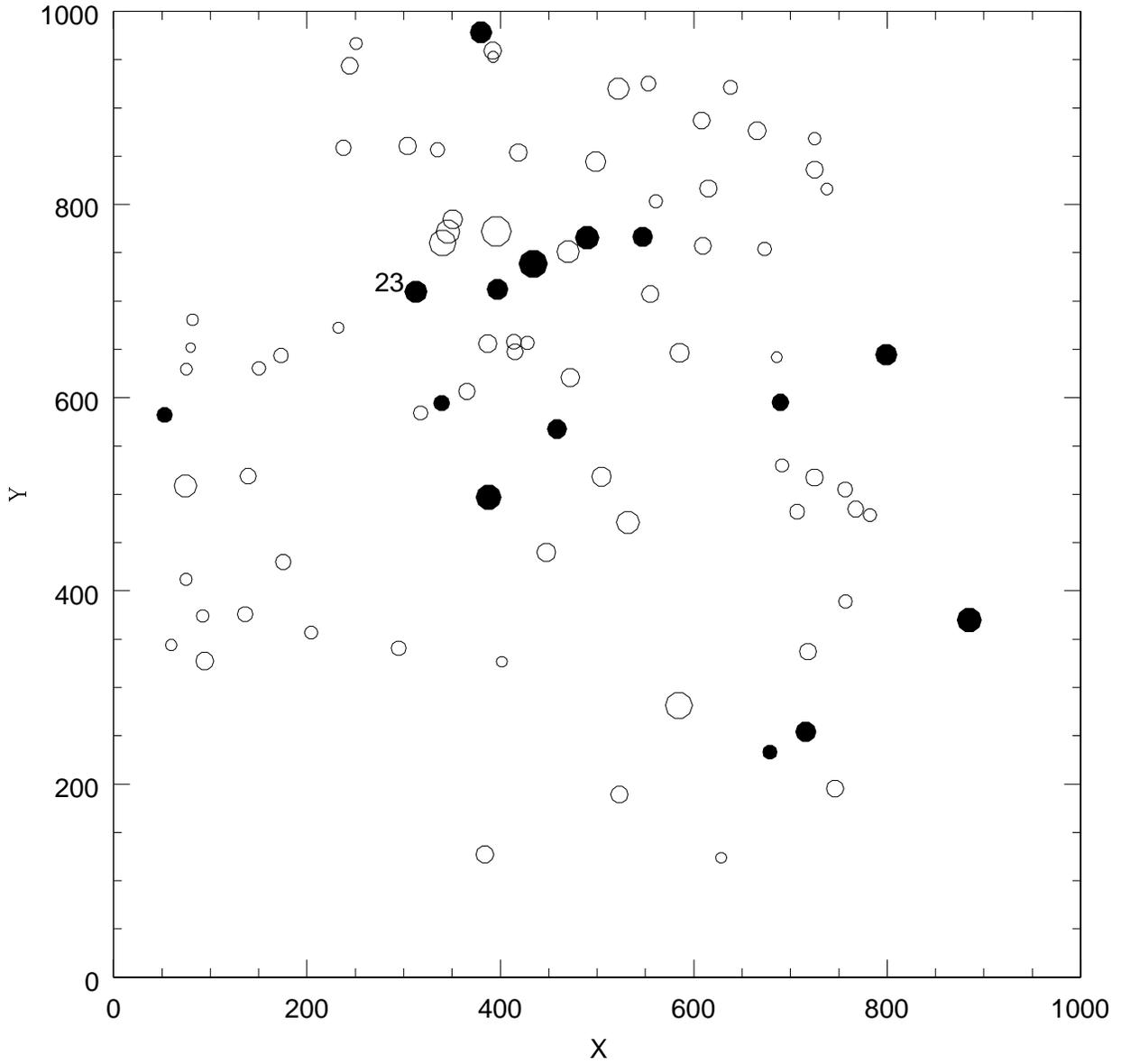}}
\caption{Cluster field of IC 4996 showing pre-MS stars 
and normal stars separately. Normal stars are shown as
open circles, pre-MS stars are shown as filled circles and
the emission stars are shown as labeled filled circles.}
\end{figure*}
\newpage
\begin{figure*}
\epsfxsize=17truecm
\centerline{\epsffile{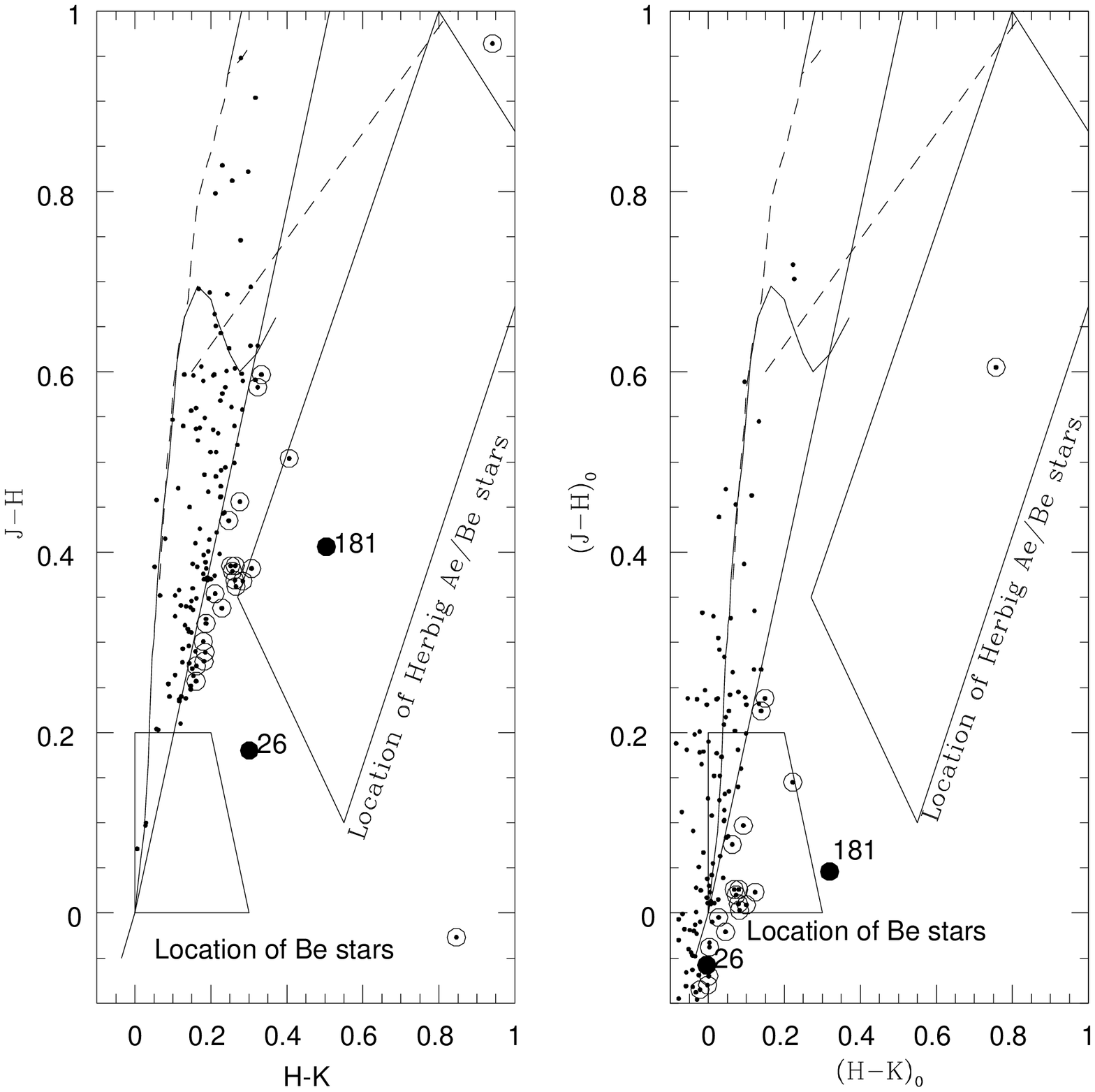}}
\caption{NIR colour-colour diagram for the cluster NGC 6910. 
Symbols and panels have the same meaning as in figure 1.}
\end{figure*}
\newpage
\begin{figure*}
\epsfxsize=17truecm
\centerline{\epsffile{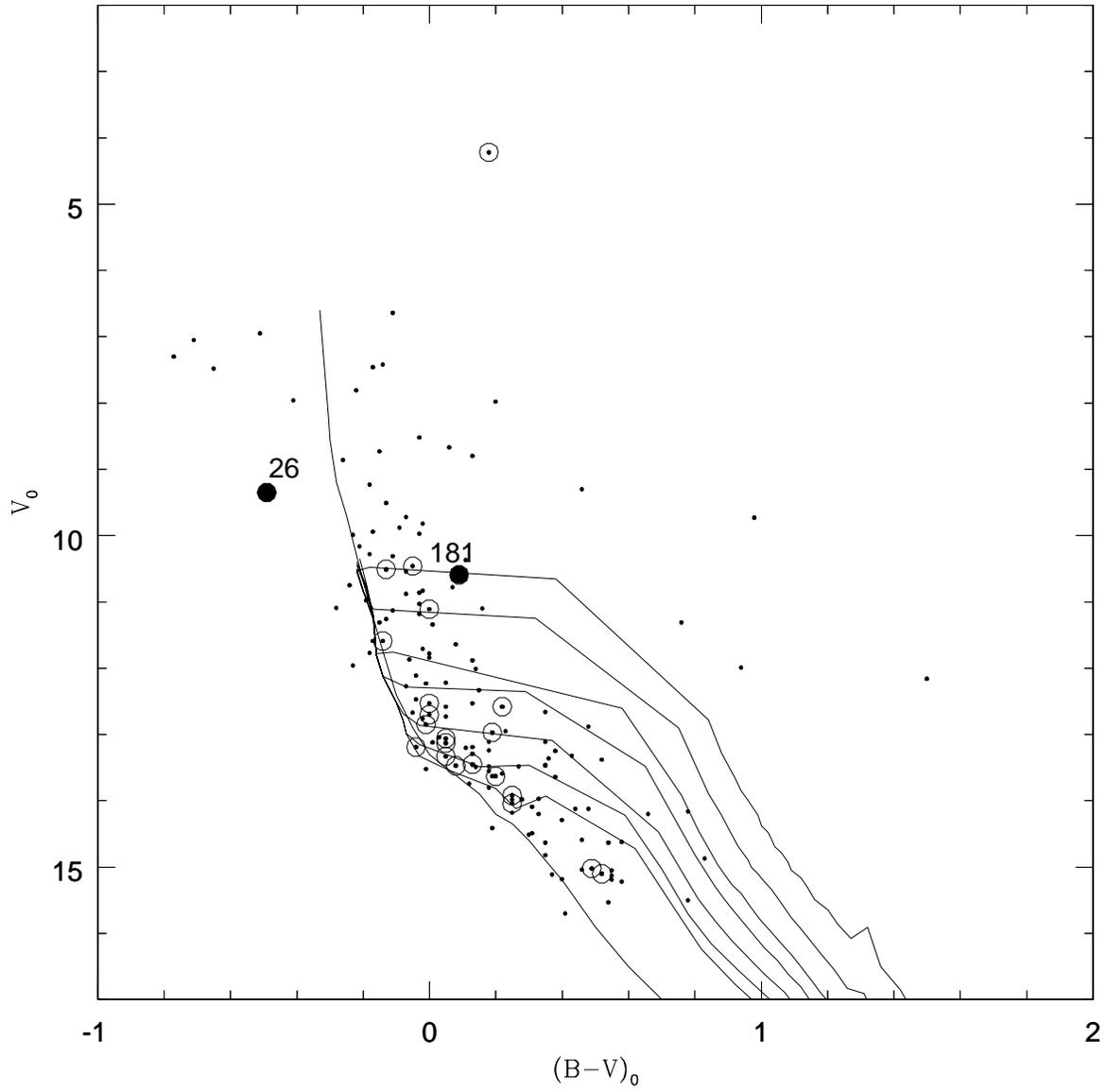}}
\caption{Pre-MS isochrone fittings for the cluster NGC 6910. The isochrones
for the ages 0.5, 1, 2, 3, 5, 7.5 and 10 Myr are shown.}
\end{figure*}
  
\newpage
\begin{figure*}
\epsfxsize=17truecm
\centerline{\epsffile{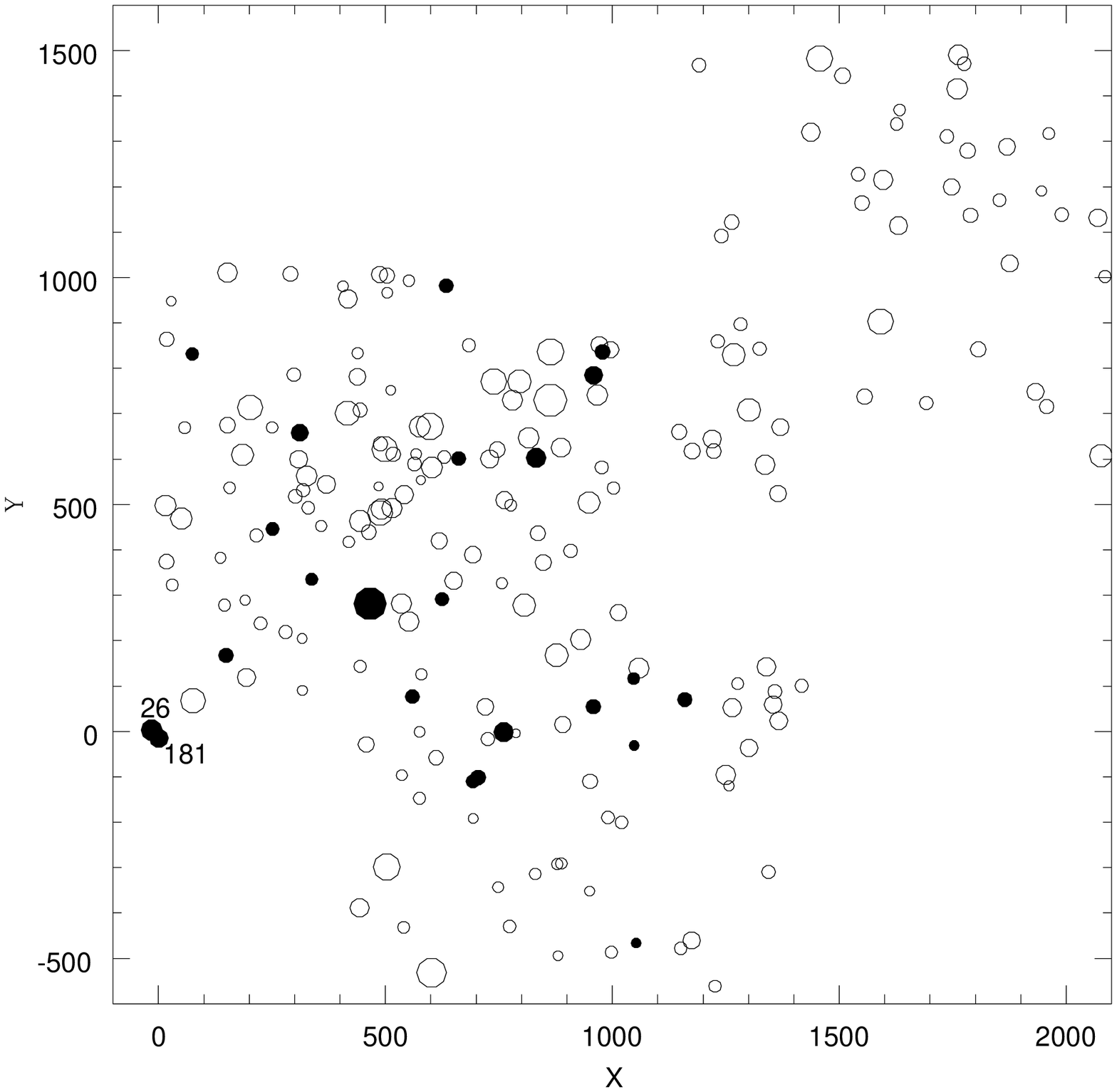}}
\caption{Cluster field of NGC 6910 showing pre-MS stars 
and normal stars separately. 
The symbols have the same meaning as in figure 3.}
\end{figure*}
\newpage
\begin{figure*}
\epsfxsize=17truecm
\centerline{\epsffile{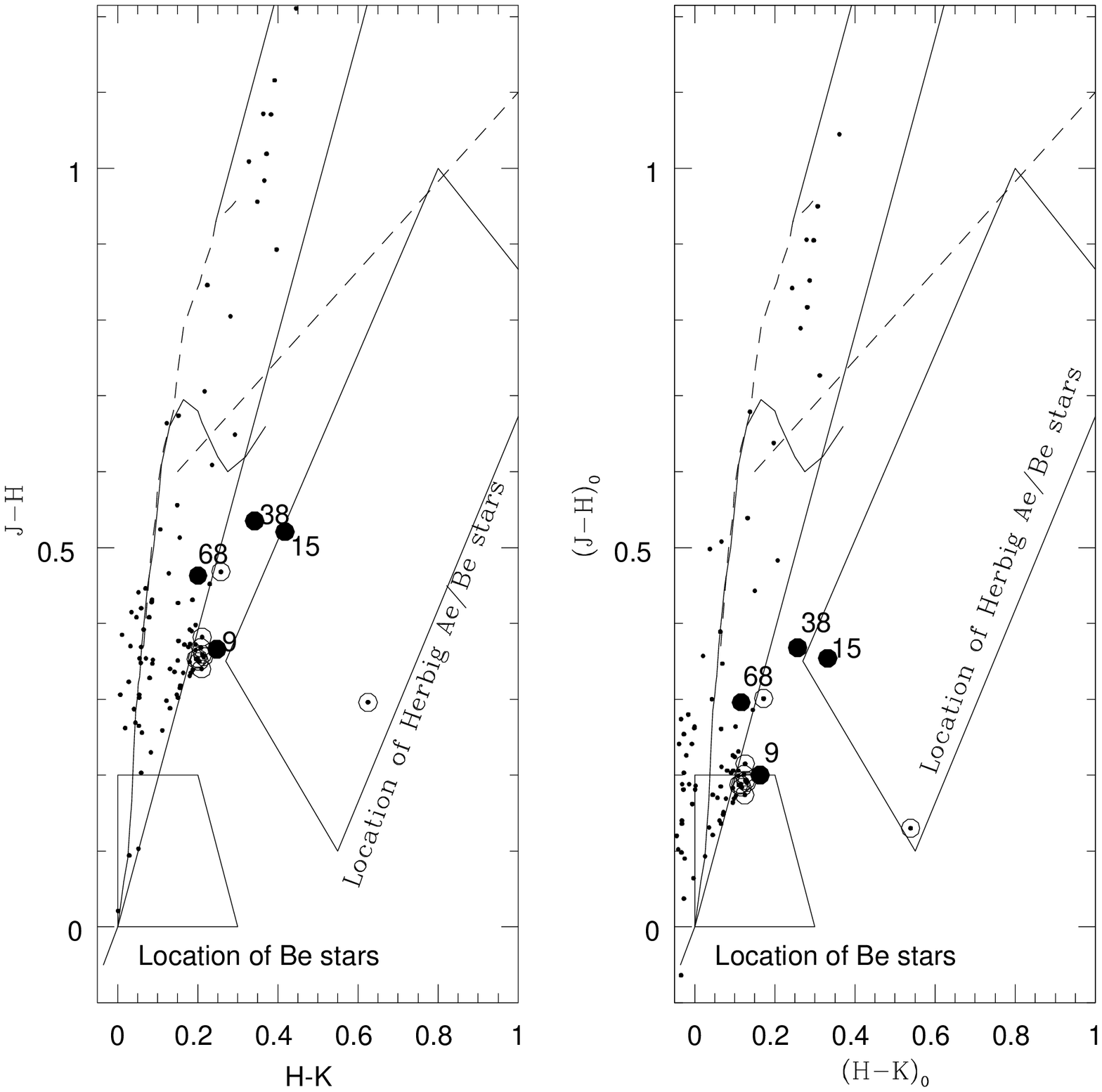}}
\caption{NIR colour-colour diagram of the cluster, Be 87.
Symbols and panels have the same meaning as in figure 1.}
\end{figure*}
\newpage

\begin{figure*}
\epsfxsize=17truecm
\centerline{\epsffile{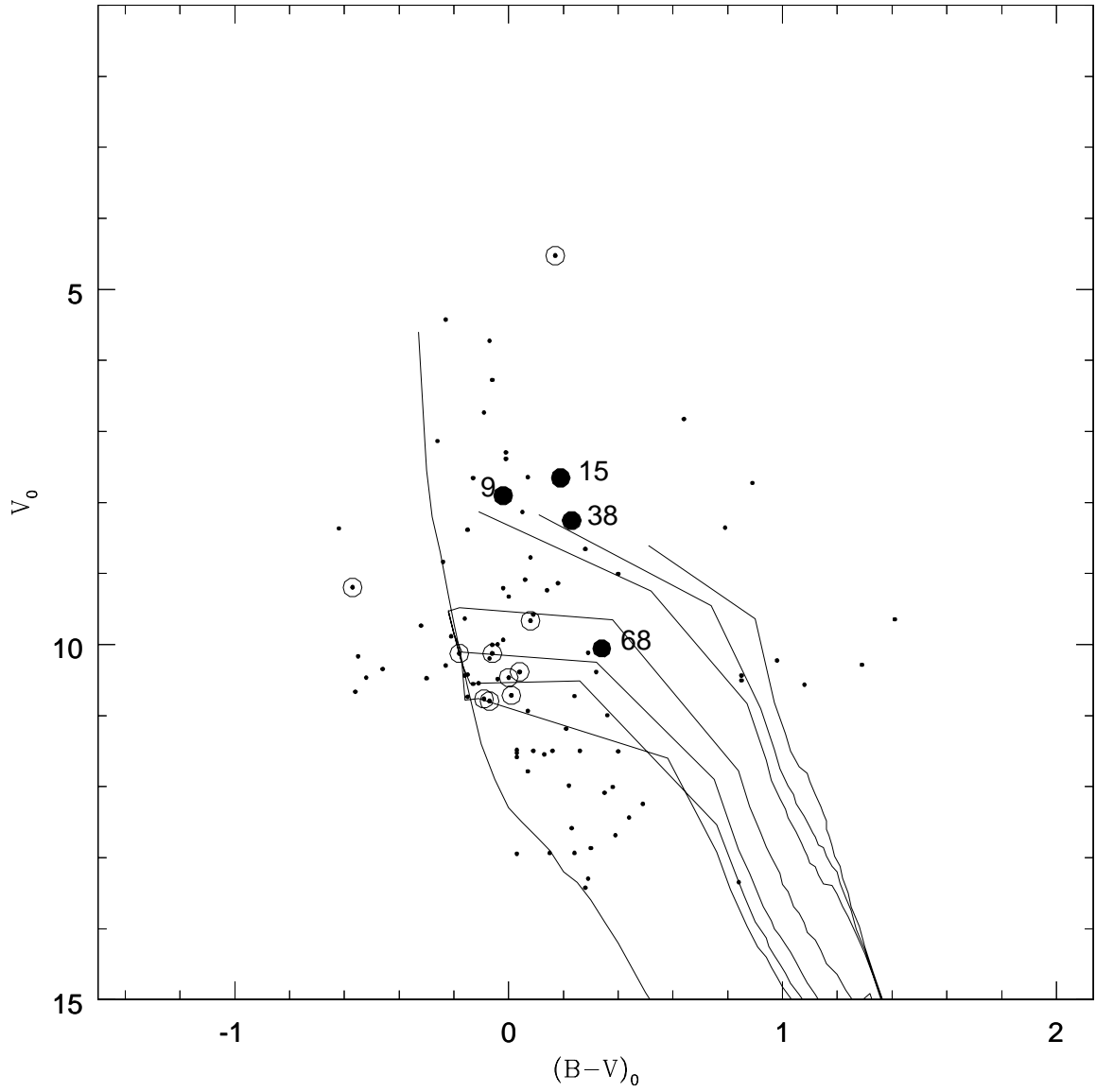}}
\caption{Pre-MS isochrone fittings for the CMD of the cluster Be 87. The isochrones
for the ages, 0.15, 0.2, 0.25, 0.5, 1, 1.5 and 2 Myr are shown.}
\end{figure*}
\newpage
\begin{figure*}
\epsfxsize=17truecm
\centerline{\epsffile{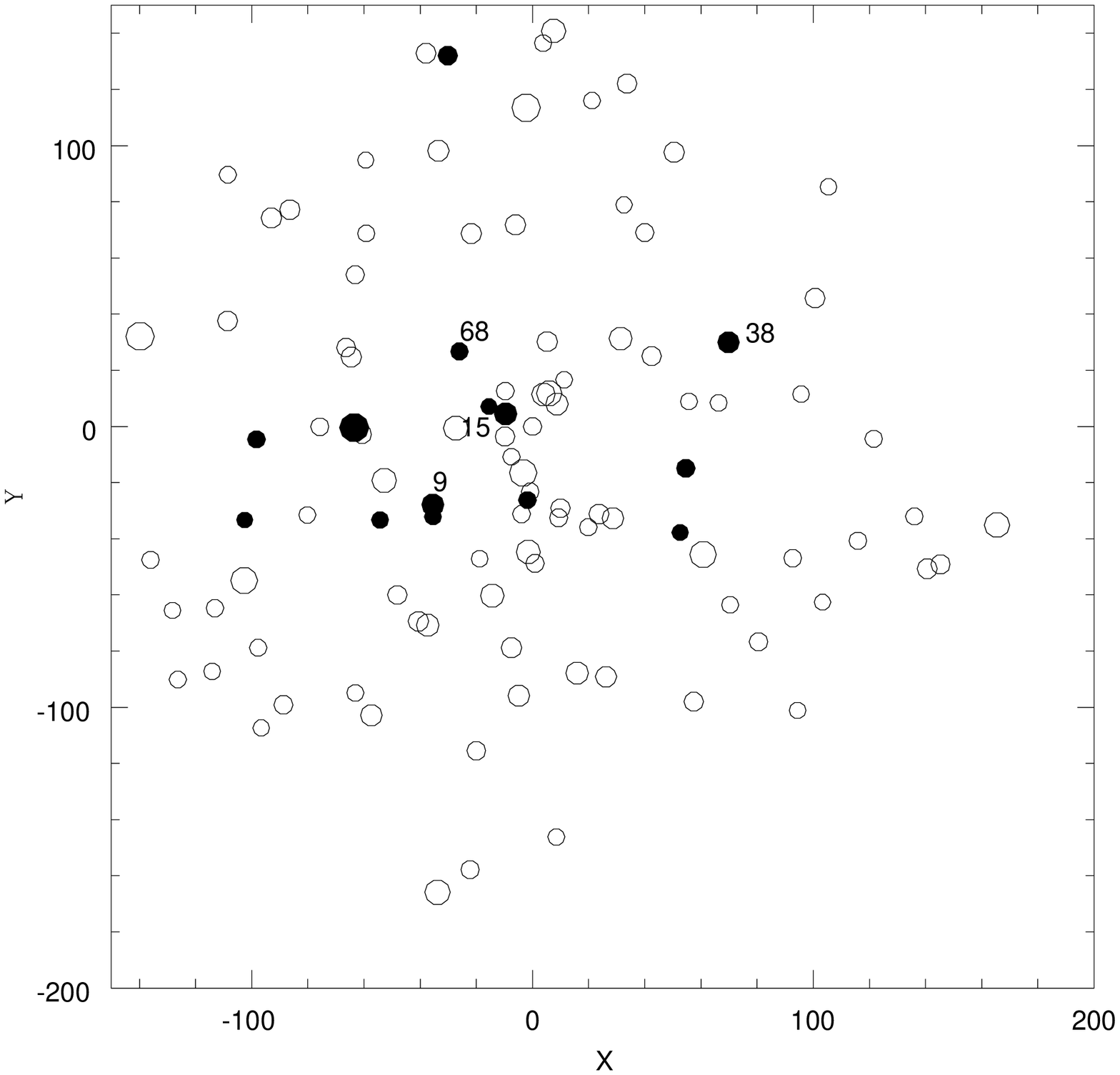}}
\caption{Cluster field of Be87 showing pre-MS stars 
and normal stars separately.
The symbols have the same meaning as in figure 3.}
\end{figure*}
\newpage
\begin{figure*}
\epsfxsize=17truecm
\centerline{\epsffile{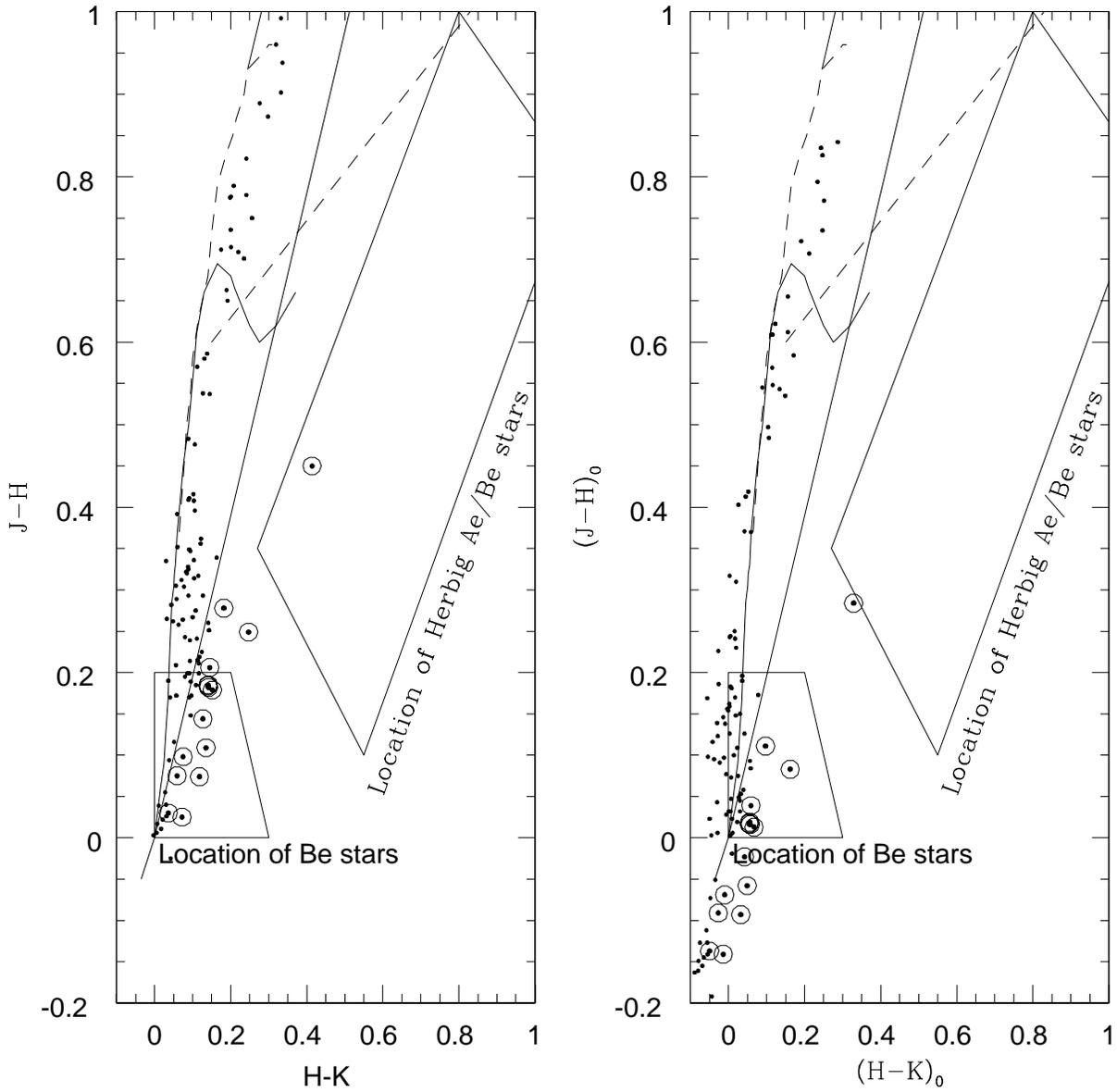}}
\caption{NIR colour-colour diagram for the cluster Biurakan 2.
Symbols and panels have the same meaning as in figure 1.}
\end{figure*}
\newpage
\begin{figure*}
\epsfxsize=17truecm
\centerline{\epsffile{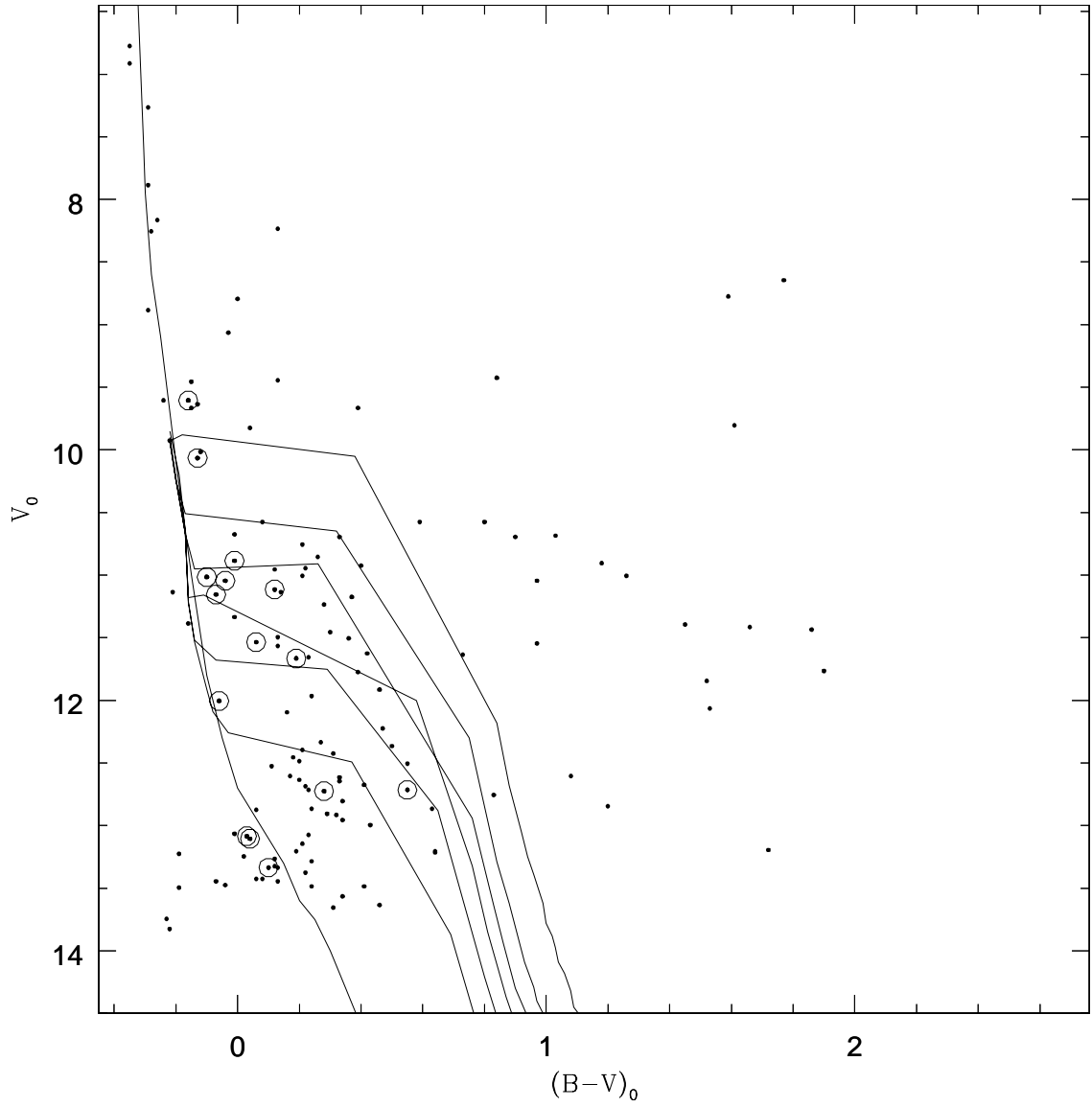}}
\caption{Pre-MS isochrone fittings for the cluster Biurakan 2. Isochrones for ages
0.5, 1, 1.5, 2, 3 and 5 Myr are shown.}
\end{figure*}
\newpage
\begin{figure*}
\epsfxsize=17truecm
\centerline{\epsffile{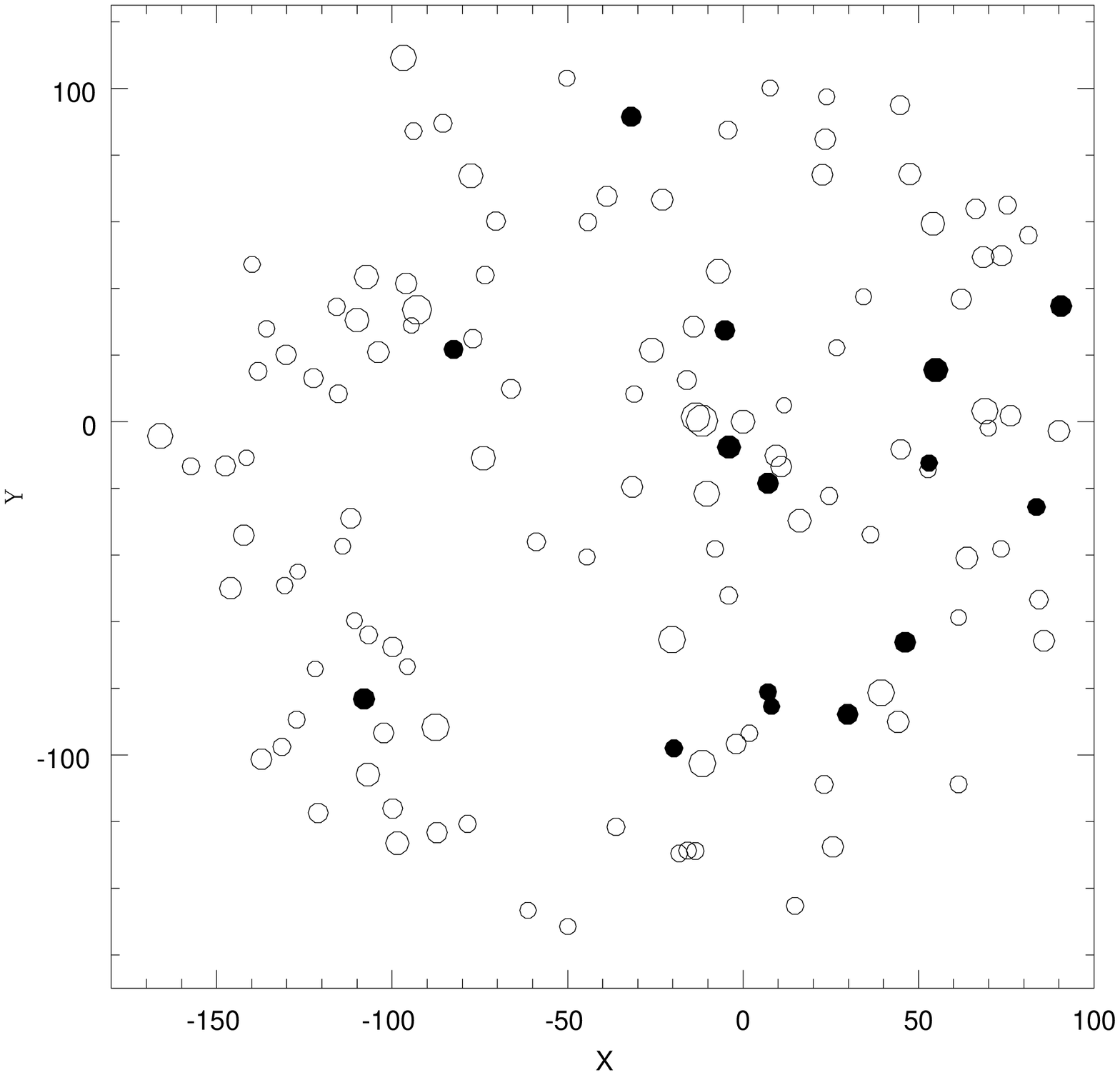}}
\caption{Cluster field of Biurakan 2 showing pre-MS stars 
and normal stars separately.
The symbols have the same meaning as in figure 3.}
\end{figure*}
\newpage
\begin{figure*}
\epsfxsize=17truecm
\centerline{\epsffile{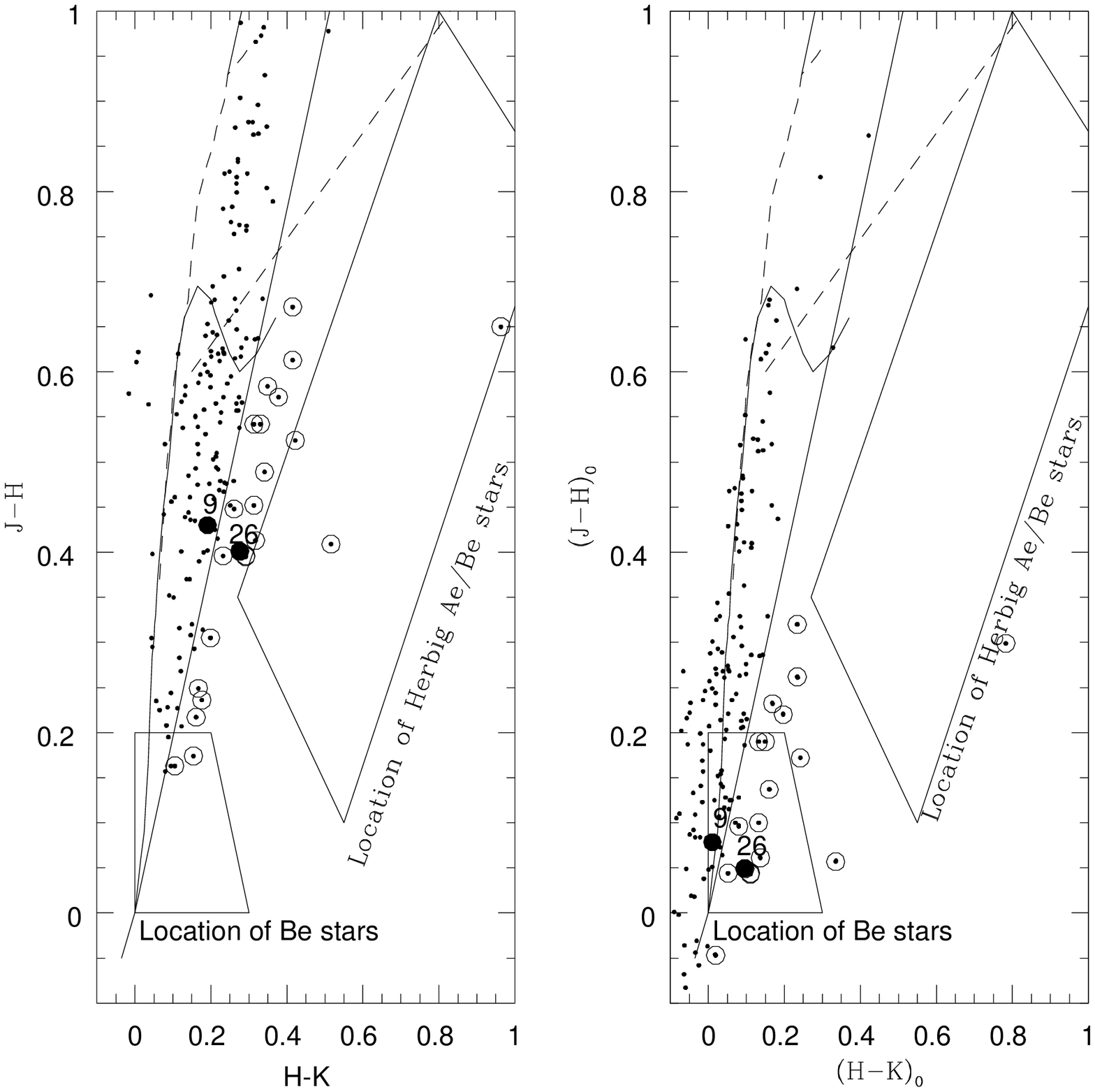}}
\caption{NIR colour-colour diagram for the cluster Be 86.
Symbols and panels have the same meaning as in figure 1.}
\end{figure*}
\newpage
\begin{figure*}
\epsfxsize=17truecm
\centerline{\epsffile{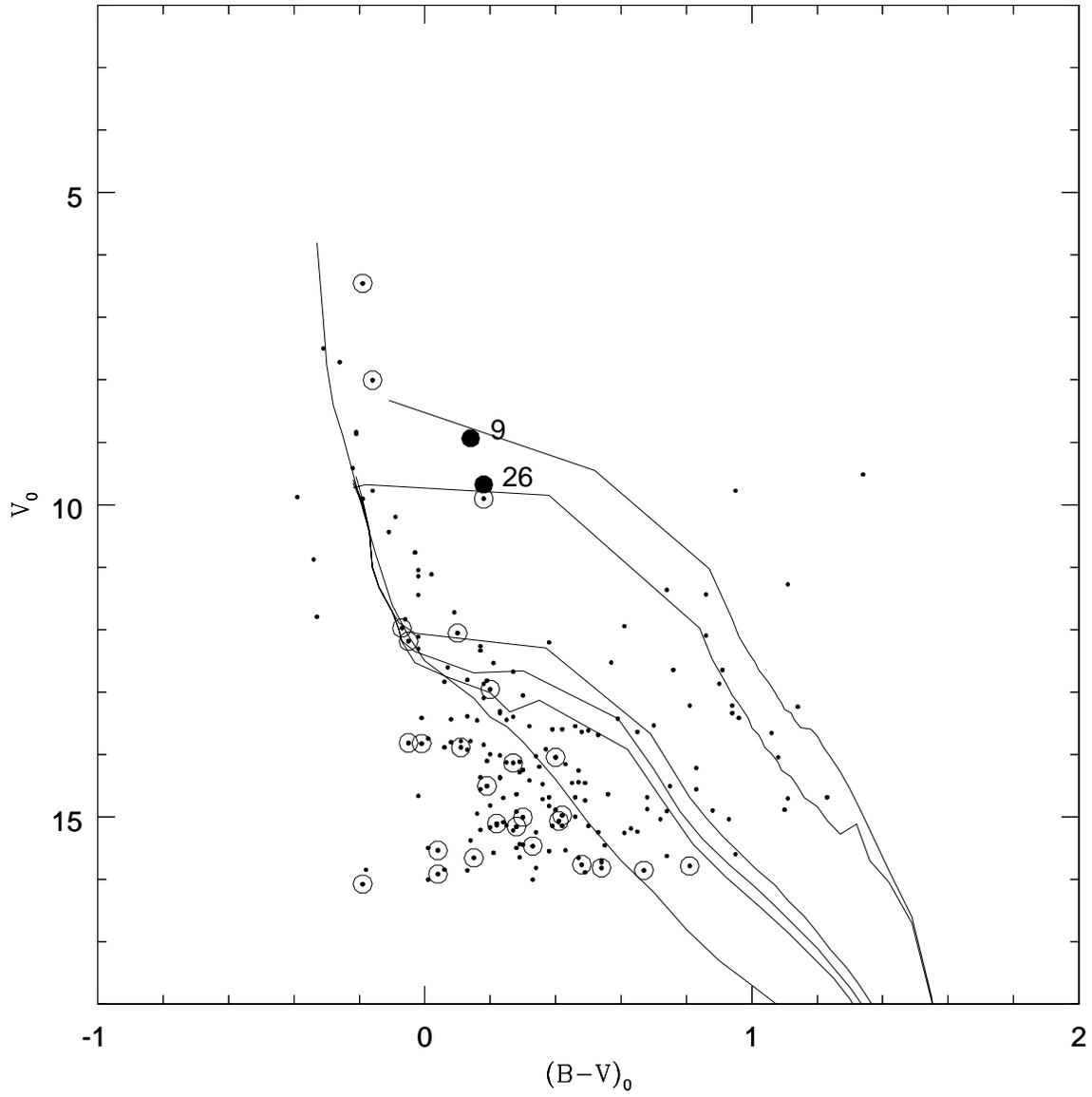}}
\caption{Pre-MS isochrone fittings for the cluster Be 86. The isochrones
for the ages, 0.25, 0.5, 5, 7.5 and 10 Myr are shown.}
\end{figure*}
\newpage
\begin{figure*}
\epsfxsize=17truecm
\centerline{\epsffile{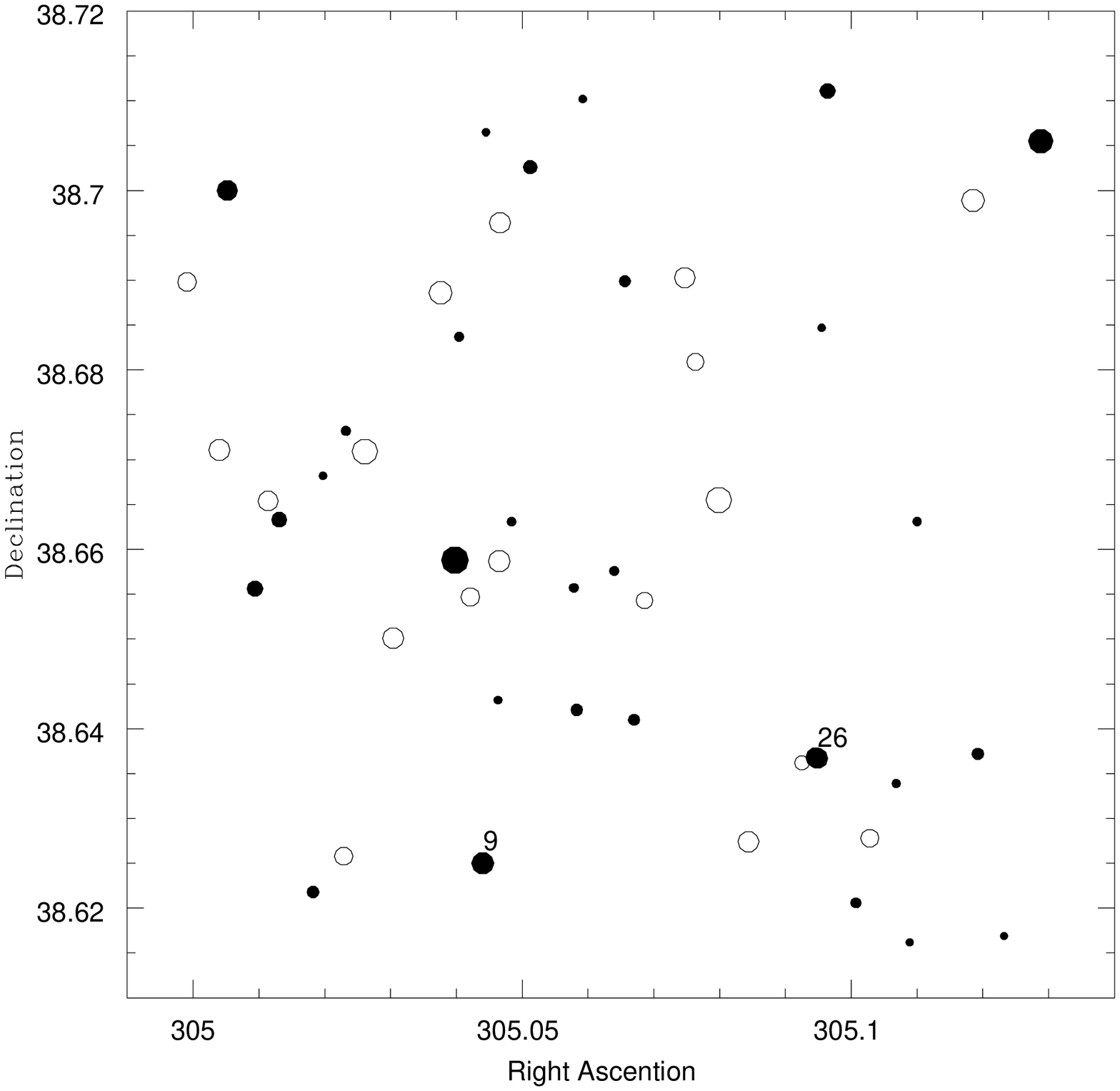}}
\caption{Cluster field of Be 86 showing pre-MS stars 
and normal stars separately.
The symbols have the same meaning as in figure 3.}
\end{figure*}

\newpage
\begin{figure*}
\epsfxsize=17truecm
\centerline{\epsffile{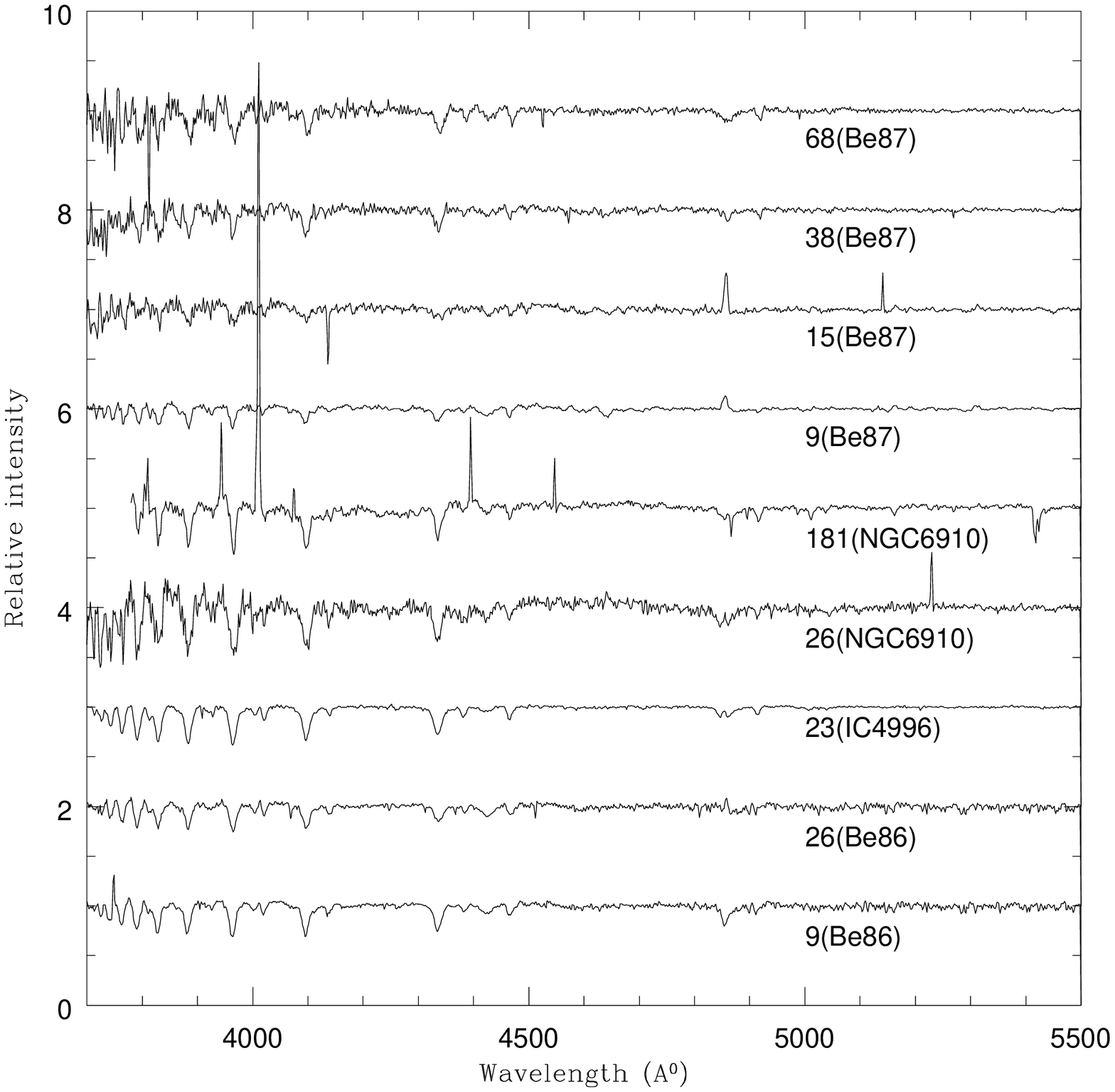}}
\caption{Flux calibrated, Normalised,Continuum fitted spectra of 9 Be stars 
in blue region.}
\end{figure*}

\newpage
\begin{figure*}
\epsfxsize=17truecm
\centerline{\epsffile{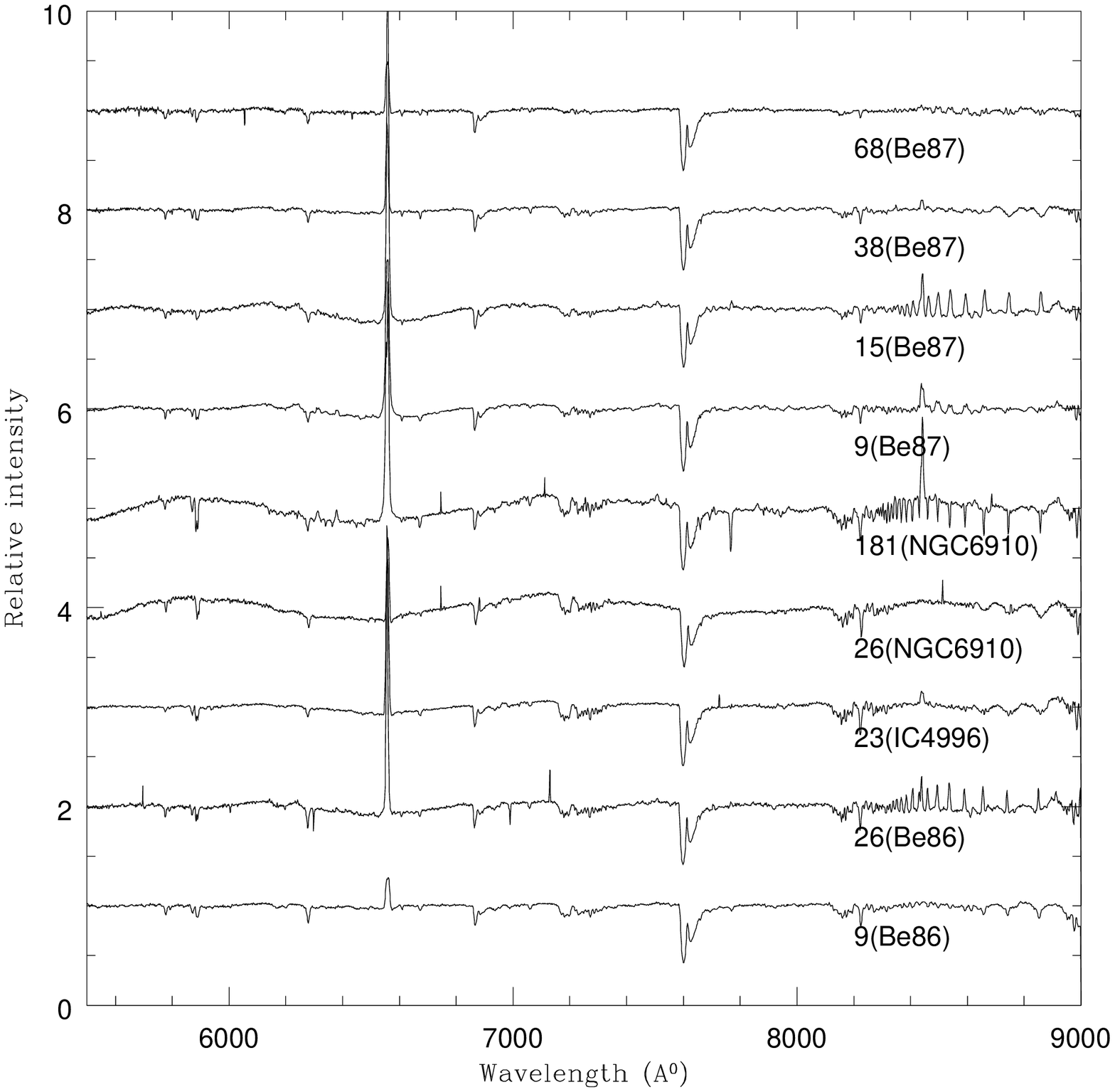}}
\caption{Flux calibrated, Normalised,Continuum fitted spectra of 9 Be stars
 in red region}
\end{figure*}

\newpage
\begin{figure*}
\epsfxsize=17truecm
\centerline{\epsffile{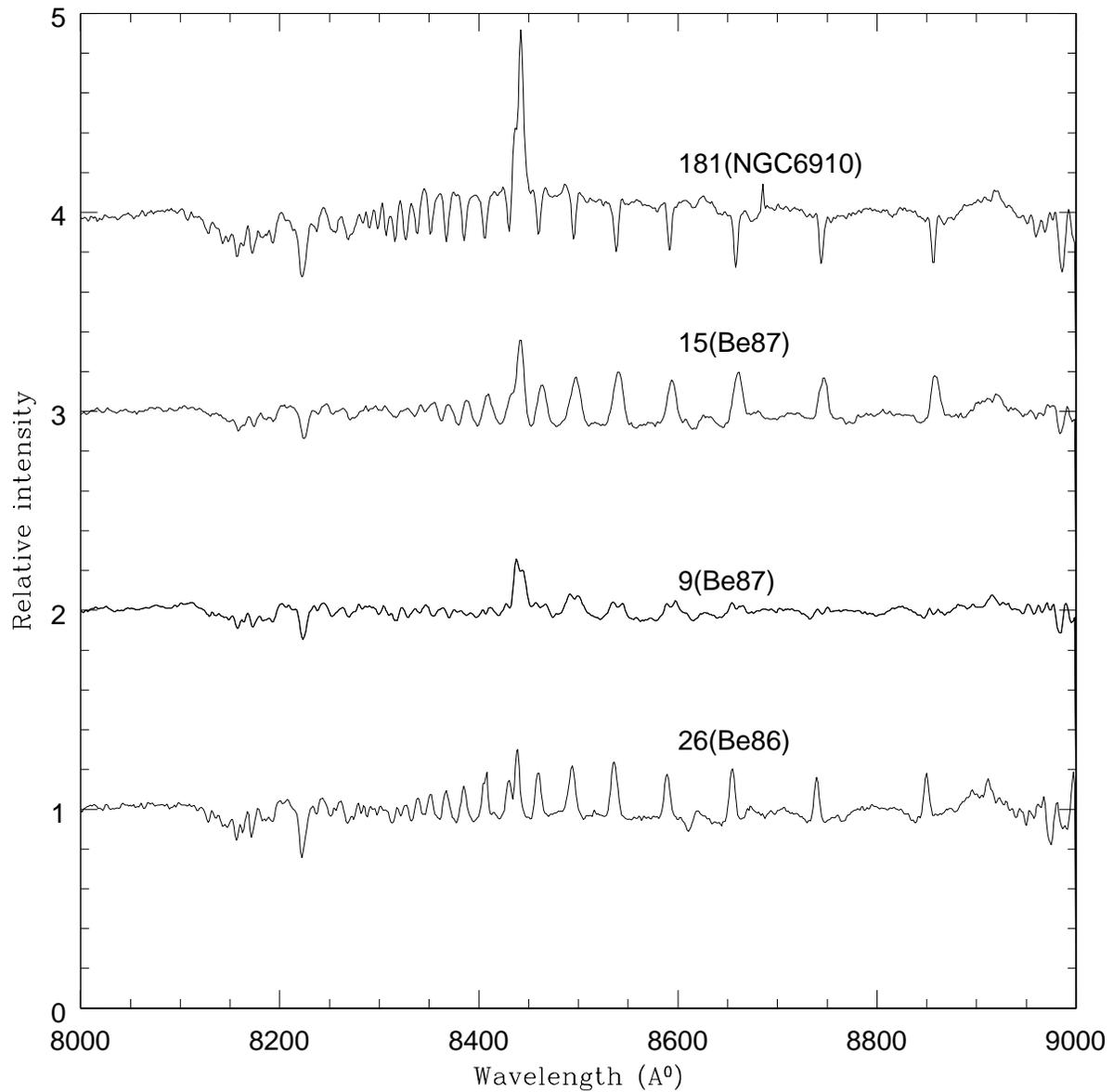}}
\caption{Spectra showing 3 stars with emission lines and 1 star with absorption
lines in red end of the optical spectrum.}
\end{figure*}

\newpage
\begin{figure*}
\epsfxsize=16truecm
\centerline{\epsffile{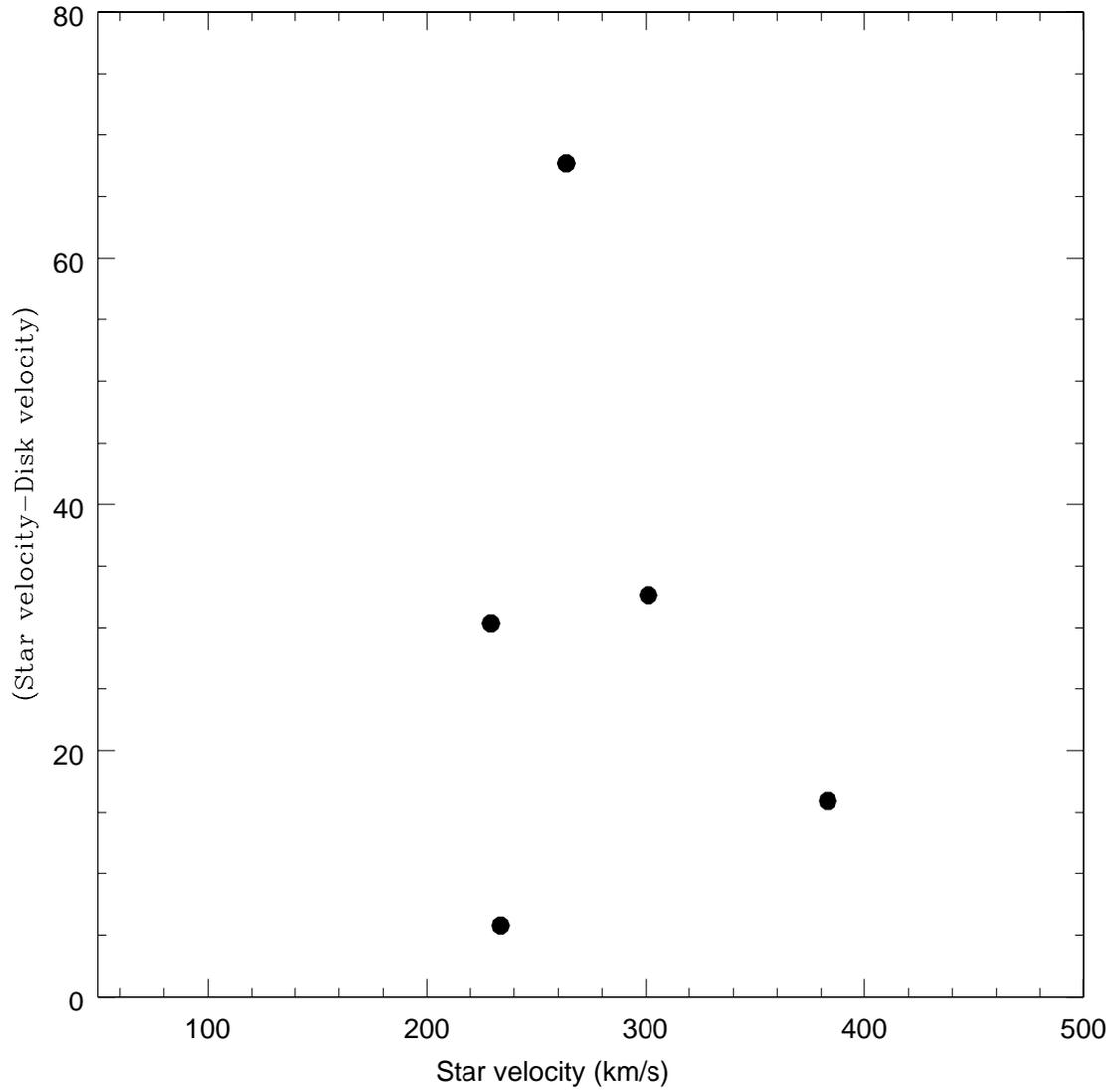}}
\caption{The relative rotational velocity between the star and its disk
is plotted against the rotational velocity of the star. The velocities
are plotted for the emission stars Be 87(38), NGC 6910 (26), Be 87(15),
IC 4996(23) and Be 86(9), in the order of increasing stellar rotational velocity.}
\end{figure*}

%

\end{document}